\documentclass[fleqn,usenatbib]{aa}  
\usepackage{graphicx}
\usepackage{txfonts}
\usepackage{multicol}
\usepackage{color,ulem}
\usepackage{amsmath, epsfig,natbib}
\usepackage{color,ulem}
\usepackage{float}
\usepackage{newtxtext,newtxmath}
\definecolor{webgreen}{rgb}{0,.5,0}
\definecolor{webbrown}{rgb}{.6,0,0}
\usepackage[colorlinks=true,hyperfootnotes=false,%
   breaklinks=true,%
   plainpages=false, bookmarksnumbered, bookmarksopen=true,
  bookmarksopenlevel=1,%
   urlcolor=webbrown, linkcolor=blue, citecolor=blue]{hyperref}
\newcommand{\kms}{\mbox{$\>{\rm km\, s^{-1}}$}}
\newcommand{\pc}{\>{\rm pc}}
\newcommand{\kpc}{\mbox{$\>{\rm kpc}$}} 
\newcommand{\kmsk}{\mbox{$\>{\rm km\, s^{-1}\, kpc^{-1}}$}}
\newcommand{\Gyr}{\mbox{$\>{\rm Gyr}$}}

\newcommand\degrees{^\circ}
\newcommand{\avg}[1]{\mbox{$\left<{#1}\right>$}}

\newcommand\gaia{{\it Gaia}}

\begin{document} 

   \title{Quadrupole signature as a kinematic diagnostic to constrain bar properties: implications for the Milky Way}
\titlerunning{bar properties from quadrupole signature}
   \author{
   Soumavo Ghosh\inst{1,2} \thanks{E-mail: soumavo@iiti.ac.in},
   Taavet Kalda \inst{2,3},
    Paola Di Matteo \inst{4},
    Gregory M. Green \inst{2},
    Sergey Khoperskov \inst{5},
   David Katz \inst{4},
   Misha Haywood \inst{4}
   }
\authorrunning{S. Ghosh et al.}

    \institute{Department of Astronomy, Astrophysics and Space Engineering, Indian Institute of Technology Indore, India - 453552
    \and
     Max-Planck-Institut f\"{u}r Astronomie, K\"{o}nigstuhl 17, D-69117 Heidelberg, Germany
   \and
   Heidelberg University, Grabengasse 1, 69117 Heidelberg, Germany
   \and 
       LIRA, Observatoire de Paris, Université PSL, Sorbonne Université, Université Paris Cité, CY Cergy Paris Université, CNRS, 92190 Meudon, France
     \and 
    Leibniz-Institut f\"{u}r Astrophysik Potsdam (AIP), An der Sternwarte 16, 14482 Potsdam, Germany
      }
   
   \date{Received XXX; accepted 22/09/2025}

  \abstract
  {The presence of a `butterfly' or a quadrupole structure in the stellar mean radial velocity ($\avg{V_R}$) field of the Milky Way is well known from the \gaia\ and the APOGEE surveys. Past studies indicated that a stellar bar can excite such a quadrupole feature in the $\avg{V_R}$ distribution. However, a systematic study investigating the co-evolution of bar and quadrupole structure is largely missing. Furthermore, whether this quadrupole structure in $\avg{V_R}$ can be used as a robust kinematic diagnostic to constrain bar properties, particularly for the Milky Way, is still beyond our grasp. Here, we investigate the bar-induced quadrupole feature using a suite of isolated $N$-body models forming prominent bars and a sample of Milky Way-like barred galaxies from the TNG50 cosmological simulation.
  We demonstrate that the properties of the quadrupole (strength, length, and orientation) are strongly correlated with the bar properties, regardless of the choice of the thin/thick disc stars; thereby making the quadrupole feature an excellent kinematic diagnostic for constraining the bar properties. In presence of spirals, the estimator which takes into account the phase-angle of $m = 4$ Fourier moment, serves as a more appropriate estimator for measuring the length of the quadrupole.
  Further, we constructed a novel \gaia-like mock dataset from a simulated bar model while incorporating the dust extinction and the broad trends of observational errors of the \gaia\ survey. The quadrupole properties (strength and length) estimated from those \gaia-like mock data are larger ($\sim 35-45$ percent) when compared with their true values. We showed that the majority of this effect is due to the uncertainty in parallax measurement. This demonstrates that the quadrupole structure in \gaia\ data is likely a result of dominant \gaia\ parallax errors/biases, almost masking the true inherent signature of the MW bar.
  }

   \maketitle
%

\section{Introduction}
\label{sec:Intro}

It is well known that the Milky Way (MW) harbours a stellar bar in the central region \citep[e.g.][]{LisztandBurton1980,Binneyetal1991,Weinberg1992,Binneyetal1997,BlitzandSpergel1991,Hammersleyetal2000,WegandGerhard2013}. However, even after dedicated efforts in the past, the properties of the MW's bar still remain ill constrained. The European Space Agency's \gaia\ mission has provided an unprecedented, holistic view of the MW by measuring the 6-D position-velocity and chemistry of $\sim 33$ million stars in the Solar Neighbourhood and beyond \citep{Katzetal2018,Drimmeletal2023}.  
However, the extinction due to dust has obscured our view towards the Galactic centre, especially at lower latitudes, that is, closer to the Galactic mid-plane \citep[see e.g.][]{Natafetal2013}. This, in turn, prevents us from identifying the exact spatial extent of the bar from the stellar density distributions (unlike the external barred galaxies) as well as measuring the properties of the MW's bar, solely based on the \gaia\ observations. 
\par
The salient properties of the MW's bar, which are of importance in the field of Galactic dynamics, are the strength, length, and the pattern speed ($\Omega_{\rm bar}$). The importance of accurately measuring these properties extends beyond the usual notion of comparing MW's bar with other external barred galaxies. Bars are known to play a pivotal role in driving the secular evolution of disc galaxies \citep[e.g. see][and references therein]{Veraetal2016}. Bars can redistribute stars and reshape metallicity distributions by radial migration \citep[e.g.][]{DiMatteoetal2013,Kubryketal2013,Halleetal2015,Khoperskovetal2020,Haywoodetal2024}, excite dark gaps along the bar minor axis \citep[e.g. see][]{Kimetal2016,Ghosh2024}, drive vertical breathing \citep{Monarietal2015,Khachaturyantsetal2022} and bending motions \citep[e.g.][]{Khoperskovetal2019}, excite ridge-like features in the phase-space \citep[e.g. see][]{Dehnen2000,Fragkoudietal2019,Tricketal2021}, impacting the structure of stellar streams in the halo \citep[e.g. see][]{Price-Whelanetal2016,Erkaletal2017,Bonacaetal2020}, funnelling gas in the inner region of galaxies; thus facilitating in starbursts and formation of nuclear discs \citep[e.g.][]{Shlosmanetal1990,Shethetal2005}, and produce large-scale streaming motions in both stars and gas \citep[e.g. see][]{SellwoodandWilkinson1993,Athanassoula1992a,Athanassoula1992b}. Furthermore, a recent study by \citet{Ghosetal2023a} showed that the accuracy of recovering the underlying (axisymmetric) potential and the distribution function (DF) in a Milky Way-like barred galaxy critically depends on the location of the survey volume with respect to the bar \citep[also see][]{Khoperskovetal2024}. Therefore, an accurate measurement of the properties of the bar in the MW is the need of the hour to quantify the bar-driven secular evolution in the MW over cosmic time. 
\par
While the jury is still out on the exact values of the bar properties in the MW, in the past, several efforts have been made towards measuring the pattern speed, length, and orientation of the MW's bar with respect to the Sun. Earlier bar pattern speed measurements, using techniques ranging from applying the Tremaine-Weinberg method \citep{TremaineandWeinberg1984} to the stellar velocity field to matching bar resonance features in the stellar velocity field to matching gas dynamics in presence of a stellar bar, favoured a fast-short bar scenario with a somewhat larger value of the pattern speed, ranging from $\sim 50$ to $60 \kmsk$ \citep[e.g., see][]{Fux1999,Dehnen2000,Debattistaetal2002,Bissantzetal2003,Antojaetal2014}. However, more recent observational measurements favour a relatively lower value for the bar pattern speed, mostly converging towards $\sim 40 \kmsk$ \citep[e.g., see][]{Sormanietal2015,Lietal2016,Portailetal2017,Bovyetal2019,Sandersetal2019,Clarkeetal2022,Lietal2022,Luceyetal2022}. As for the bar length of the MW, earlier study by \citet{Hammersleyetal1994}, using the star counts from the Two-Micron Galactic Survey, estimated the bar length to be $\sim 4 \kpc$. In addition, \citet{Weggetal2015}, using the red clump giant (RCG) stars, measured the MW's bar length to be $\sim 5 \kpc$. However, recent study by \citet{Luceyetal2022}, employing a technique based on the maximal extent of trapped bar orbits as an estimate of bar length, estimated the bar length to be $\sim 3.5 \kpc$. While this latter method relies on robust dynamical arguments, the resulting bar length measurement critically depends on the assumed underlying potential of the Galaxy. Furthermore, the bar length, computed from the stellar density field, can be overestimated if the bar is connected to the spiral structure \citep[e.g. see][]{Hilmietal2020,GhoshandDiMatteo2024,Visloskyetal2024}. As for the orientation of the MW's bar, \citet{Weggetal2015} estimated that the bar is at $\sim 28^{\degrees} - 33^{\degrees}$ with respect to the Sun. The bar strength and shape are even less well known \citep[however, see the models to compare the strength of the bar resonances in the Solar neighbourhood from the \gaia\ data in][]{Monarietal2019}.
\par
From the \gaia\ Data Release 3 (hereafter \gaia\ DR3), \citet{Drimmeletal2023} showed the existence of a `quadrupole' or butterfly-like pattern in the stellar mean radial velocity ($\avg{V_R}$) field of red giant branch (RGB) stars (see their Fig. 16). This feature has been reported previously from the APOGEE line-of-sight velocities and the \gaia\ DR2 astrometry \citep{Bovyetal2019} and  from the APOGEE line-of-sight velocities and the \gaia\ EDR3 astrometry \citep{Queirozetal2021}. Initial theoretical studies, by means of test particle simulation as well as by using self-consistent $N$-body simulation of a barred galaxy, showed that a quadrupole feature in the stellar mean radial velocity field ($\avg{V_R}$) is excited by a central stellar bar \citep[e.g. see][]{Bovyetal2019,Drimmeletal2023}. Furthermore, a recent study by \citet{Visloskyetal2024}, by empirically matching the maps of $\avg{V_R}$ from the \gaia\ DR3 and the barred-spiral galaxies from the TNG50 cosmological simulation, proposed that the MW's stellar velocity field is consistent with a short bar (with a bar length  $\sim 3.6 \kpc$) connected to a spiral arm. While it is theoretically understood that a bar can excite a quadrupole feature in the stellar mean radial velocity field, it still remains beyond our grasp whether one can use the quadrupole feature as a robust kinematic diagnostic to constrain (some of) the properties of the bar.  A directly related question would be whether one can use the quadrupole feature, as seen from the \gaia\ data, to put stringent constraints on the bar properties of the MW from a purely dynamical argument. If yes, it can potentially mitigate the existing conundrum of measuring the properties of the bar in the MW from the stellar density field (which is severely affected by the dust attenuation). We aim to pursue this here.
\par
In this paper, we carry out a systematic study to test the reliability and robustness of the quadrupole pattern in the stellar mean radial velocity field ($\avg{V_R}$) as a kinematic diagnostic to put constraints on the bar properties. To achieve that, we make use of a suite of collisionless $N$-body models (having both the thin and the thick discs) which forms a prominent bar and a boxy/peanut bulge. In addition, we use a sample of barred galaxies, selected from the TNG50 cosmological simulation to augment this study. Within the scope of this paper, we first quantify the properties of the quadrupole feature, and then, we systematically investigate how robustly the properties of the quadrupole trace the properties of the bar. In addition, we investigate how the biases and the errors in measuring parallax, proper motion, and radial velocity of stars, similar to the \gaia\ DR3, could influence the measured properties of the bar as inferred from the properties of the quadrupole feature. Lastly, we measure the properties of the quadrupole feature in the MW while using the full 6-D phase-space information from the \gaia\ DR3. 
\par
The rest of the paper is organised as follows. Sect.~\ref{sec:simu_setup} provides a brief description of the isolated $N$-body simulations as well the barred galaxies from the TNG50 suite of cosmological simulation, used for this study. Sect.~\ref{sec:quadrupole_investigation} provides the details of the quantification of the properties of the quadrupole feature as well as their correlation with the properties of the bar. Sect.~\ref{sec:MW_bar_properties} contains the details of the influence of \gaia\ -like uncertainties on inferring the bar properties from the properties of the quadrupole as well as contains results pertaining to the quantification of quadrupole's properties from the \gaia\ DR3.  
Sect.~\ref{sec:conclusion} summarises the main findings of this work.

\section{Simulated barred galaxies}
\label{sec:simu_setup}
Here, we briefly describe the initial equilibrium configurations and the structural properties of the suite of the isolated, collisionless $N$-body models as well as the sample of the barred galaxies, chosen from the TNG50 cosmological simulation. 

\subsection{Isolated barred models}
\label{sec:isolated_abrredmodel}

A total of 14 isolated, collisionless $N$-body models are used for this work. Below, we briefly mention their structural properties and the initial equilibrium set-up.
\par
\paragraph{thin+thick models:} A total of 13 such thin+thick models (with different disc geometry and thick disc mass fraction), taken from \citet{Ghoshetal2023a}, are considered in this work. The initial equilibrium configuration of each of these models consists of a thin and a thick stellar disc which are embedded in a live dark matter halo. Each of the thin and thick discs is modelled with a Miyamoto-Nagai profile \citep{MiyamatoandNagai1975}, having $R_{\rm d}$, $z_{\rm d}$, and $M_{\rm d}$ as the characteristic disc scale length, the scale height, and the total mass of the disc, respectively. The scale heights of the thick and thin discs are fixed to $0.9 \kpc$ and $0.3 \kpc$, respectively. The total stellar mass is fixed to to $1 \times 10^{11} M_{\odot}$ across the suite of simulations while $f_{\rm thick}$ (denoting the mass fraction in the thick disc) is varied from 0 to 0.7 in different models. The dark matter halo is modelled by a Plummer sphere \citep{Plummer1911}, having $R_{\rm H}$ ($= 10 \kpc$) and $M_{\rm dm}$ ($= 1.6 \times 10^{11} M_{\odot}$) as the characteristic scale length and the total halo mass, respectively. The values of the key structural parameters for the thin and thick discs are mentioned in Table~\ref{table:key_param}. A total of $1 \times 10^6$ particles are used to model the stellar (thin+thick) disc while a total of $5 \times 10^5$ particles are used to model  the dark matter halo.
\par
Following the iterative method algorithm by \citet{Rodionovetal2009}, the initial conditions of the discs are obtained while keeping the velocity dispersion (along the radial and vertical directions) fixed and letting the density to vary until the desired equilibrium confirmation is achieved. The simulations are run using a TreeSPH code by \citet{SemelinandCombes2002} which employs a hierarchical tree method \citep{BarnesandHut1986} with opening angle $\theta = 0.7$ to compute the gravitational forces. In addition, a Plummer potential was employed for softening the gravitational forces with a softening length $\epsilon = 150 \pc$. We evolved all the models for a total time of $9 \Gyr$. For further details, the reader is referred to  \citet{Fragkoudietal2017} and \citet{Ghoshetal2023a}. Each of the models forms a prominent stellar bar which subsequently undergoes a vertical buckling instability to form a boxy/peanut (hereafter b/p) bulge \citep[for details, see][]{Ghoshetal2023a,Ghoshetal2024}. We mention that in rthickE models, $R_{\rm d, thick} = R_{\rm d, thin}$;  in rthickS models, $R_{\rm d, thick} < R_{\rm d, thin}$; and in rthickG models, $R_{\rm d, thick} > R_{\rm d, thin}$ where $R_{\rm d, thin}$ and $R_{\rm d, thick}$ denote the scale length for the thin and thick disc, respectively. Furthermore, following the convention used in \citet{Ghoshetal2023a}, any thin+thick model is referred as a unique string `{\sc [model configuration][thick disc fraction]'} where {\sc [model configuration]} denotes the corresponding thin-to-thick disc scale length configuration while {\sc [thick disc fraction]} denotes the value of $f_{\rm thick}$.

\paragraph{sim6 model:} This is a higher resolution collisionless $N$-body model (as compared to other thin+thick models used here) which also forms a prominent bar and subsequently undergoes a vertical buckling instability to form a prominent b/p structure. In addition, this model has been extensively used in studying the bar-spiral driven chemo-dynamical evolution of Milky Way-like galaxies \citep[for details, see][]{Fragkoudietal2018,Fragkoudietal2019,Khoperskovetal2020a,Khoperskovetal2020b}. The initial equilibrium configuration consists of a thin disc, an intermediate disc, and a thick disc (each modelled with a Miyamoto-Nagai profile with $R_{\rm d}$, $z_{\rm d}$, and $M_{\rm d}$ as the characteristic disc scale length, the scale height, and the total mass of the disc) and the stellar discs are embedded in a concentric live dark matter halo. The dark matter halo is modelled by a Plummer sphere, having $R_{\rm H}$ ($= 21 \kpc$) and $M_{\rm dm}$ ($= 3.7 \times 10^{11} M_{\odot}$) as the characteristic scale length and the total halo mass, respectively. The scale lengths of the thin, the intermediate, and the thick disc are set to $4.8 \kpc$, $2 \kpc$, and $2 \kpc$, respectively. The total stellar mass (thin+intermediate+thick) of the model is fixed to $\sim 8.7 \times 10^{10} M_{\odot}$ where the thin disc contributes 50 percent of the total stellar mass while the intermediate and the thick disc constitutes 30 percent and 20 percent of the total stellar mass, respectively. The scale heights of the thin, intermediate, and thick disc are set to $0.15 \kpc$, $0.3 \kpc$, and $0.6 \kpc$, respectively.  For further details, the reader is referred to \citet{Fragkoudietal2019}.
\par
The initial equilibrium configuration is achieved using the same iterative method algorithm by \citet{Rodionovetal2009} as before. The simulations are run using a parallel MPI tree-code \citep{Khoperskovetal2014} which takes into account the adaptive spatial decomposition of particle space between nodes, and with opening angle $\theta = 0.7$ to compute the gravitational forces. A Plummer potential was employed for softening the gravitational forces with a softening length $\epsilon = 50 \pc$. A total of $1.5 \times 10^7$ particles are used in the model with $ 1 \times 10^7$ used for the disc, and $5 \times 10^6$ for the dark matter halo. The model is evolved for a total time of $7 \Gyr$ \citep[for further details, see][]{Fragkoudietal2019}.
%
%
\begin{table}
\centering
\caption{Key structural parameters for the equilibrium models.}
\begin{tabular}{ccccc}
\hline
\hline
 Model$^{(1)}$ & $f_{\rm thick}$$^{(2)}$ & $R_{\rm d, thin}$$^{(3)}$ & $R_{\rm d, thick}$$^{(4)}$ \\
\\
 && (kpc) & (kpc) \\
\hline
rthick0.0 & 0 & 4.7 & - \\
rthickS0.1 & 0.1 & 4.7 &  2.3  \\
rthickE0.1 & 0.1 & 4.7 &  4.7 \\
rthickG0.1 & 0.1 & 4.7 &  5.6  \\
rthickS0.3 & 0.3 & 4.7 &  2.3  \\
rthickE0.3 & 0.3 & 4.7 &  4.7  \\
rthickG0.3 & 0.3 & 4.7 &  5.6  \\
rthickS0.5 & 0.5 & 4.7 &  2.3  \\
rthickE0.5 & 0.5 & 4.7 &  4.7 \\
rthickG0.5 & 0.5 & 4.7 &  5.6 \\
rthickS0.7 & 0.7 & 4.7 &  2.3 \\
rthickE0.7 & 0.7 & 4.7 &  4.7 \\
rthickG0.7 & 0.7 & 4.7 &  5.6 \\
\hline
\end{tabular}
\newline{
(1) Name of the model; (2) thick disc mass fraction; (3) Scale length of the thin disc; (4) Scale length of the thick disc.}
\label{table:key_param}
\end{table}

\subsection{Barred galaxies from the TNG50 simulations}
\label{sec:TNG50_suite}

In this work, we also analyse a sample of the MW and M31 analogues~\citep{Pillepichetal2023}
selected from the TNG50 simulation~\citep{Nelsonetal2019a,Nelsonetal2019b, Pillepichetal2019}\footnote{\url{https://www.tng-project.org/data/milkyway+andromeda/}}. TNG50 is a magneto-hydrodynamical simulation of the formation and evolution of galaxies in a 51.7 comoving Mpc cube from redshift $\approx$ 127 to redshift 0. It is run with the moving-mesh code AREPO~\citep{Springeletal2010} and uses the fiducial TNG galaxy formation model~\citep{Weinbergeretal2017,Pillepichetal2018} with a mass resolution of $m_\text{baryon} = 8.5 \times 10^4 \text{M}_\odot$, $m_\text{DM} = 4.5 \times 10^5 \text{M}_\odot$; and a spatial resolution of star-forming gas of $\sim 150 \pc$ \citep{Pillepichetal2023}. 

The selection criteria of MW/M31 analogues at $z = 0$ include the following: (i) the galaxy stellar mass is in the following range: $\rm M_*(<30 kpc) = 10^{10.5-11.2}M_\odot$; (ii) a disc-like stellar morphology; (iii) no other galaxy with stellar mass $> 10^{10.5} M_\odot$ is within $500$~kpc distance; and (iv) the total mass of the halo host is smaller than that typical of massive groups $< 10^{13}M_\odot$ \citep[for further details, see][]{Pillepichetal2023}. Since in this work we are interested in barred galaxies, from the parental sample of $198$ MW and M31 analogues we select 55 galaxies with the most prominent bars.

\section{Quantifying bar properties via quadrupole signature in the mean radial velocity}
\label{sec:quadrupole_investigation}

\begin{figure*}
\centering
\resizebox{0.95\linewidth}{!}{\includegraphics{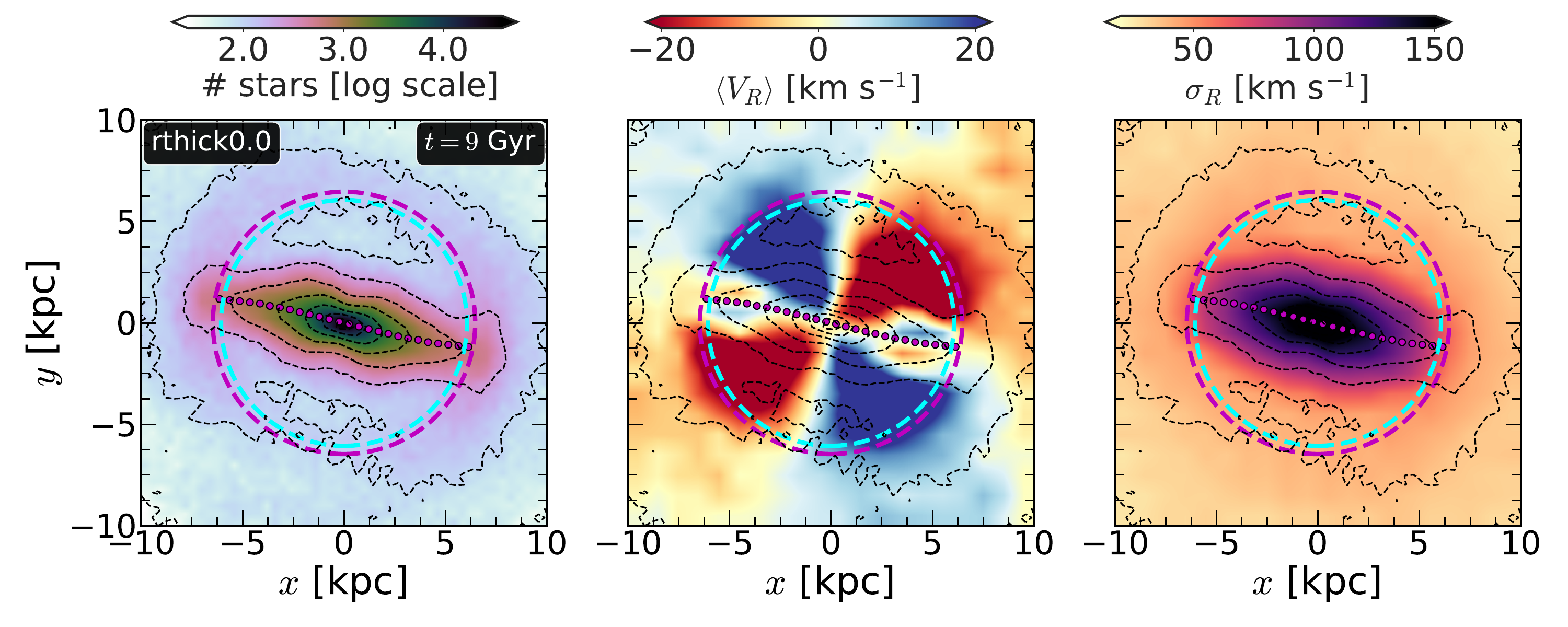}}
\caption{Tracing the bar with the quadrupole feature: distribution of stellar surface density (\textit{left panel}), mean radial velocity, $\avg{V_R}$ (\textit{middle panel}), and radial velocity dispersion, $\sigma_R$ (\textit{right panel}), calculated in the $(x-y)$-plane, for the model rthick0.0, at the end of the simulation run ($t = 9 \Gyr$). Black dashed lines denote the contours of constant surface density. The cyan dashed circle denotes the bar length, $R_{\rm bar}$, and the magenta dashed circle denotes the extent of the quadrupole feature, $R_{\rm quadrupole}$. The magenta points denote the spatial distribution of the phase-angle of the $m=4$ Fourier moment ($\varphi_4$). The bar excites a prominent quadrupole pattern in the mean radial velocity field, and the orientation of the quadrupole pattern agrees fairly accurately with the orientation of the bar. Here, we used a galactocentric cylindrical coordinate system ($R, \phi, z$) to calculate $\avg{V_R}$ and $\sigma_R$.}
\label{fig:density_maps_endstep_allmodels}
\end{figure*}

Fig.~\ref{fig:density_maps_endstep_allmodels} shows one example of the distribution of the stellar density, mean radial velocity ($\avg{V_R}$), and the radial velocity dispersion ($\sigma_R$) in the $(x-y)$-plane, calculated at $t = 9 \Gyr$ for the model rthick0.0. We used a galactocentric cylindrical coordinate system $(R, \phi, z)$, to compute the distribution of $\avg{V_R}$ and $\sigma_R$. The model harbours a prominent stellar bar in the central region, and the associated $\avg{V_R}$ map displays a prominent quadrupole feature. The model gets heated preferentially along the 2-D extent of the bar, similar to the findings of \citet{Ghoshetal2023a}. In Appendix~\ref{appen:density_and_Vr_maps}, we show the distribution of $\avg{V_R}$ in the $(x-y)$-plane, calculated at the end of the simulation ($t = 9 \Gyr$), for all the thin+thick models considered here (see Fig.~\ref{fig:Vr_maps_allmodels_endstep} there). In addition, we show the corresponding $\avg{V_R}$ maps in the $(x-y)$-plane for a sample of TNG50 barred galaxies (with varying bar morphology, see Fig.~\ref{fig:Vr_maps_TNG50_endstep} in Appendix~\ref{appen:density_and_Vr_maps}). A prominent bar is always associated with a clear quadrupole feature in the  $\avg{V_R}$ map; thereby demonstrating that the quadrupole feature in the $\avg{V_R}$ map is a (kinematic) part and parcel of the bar in the density distribution.
\par
Next, we quantify the strength, length, and the orientation of the quadrupole feature present in the face-on $\avg{V_R}$ maps. This is achieved by means of the Fourier decomposition of the mean radial velocity field. Subsequently we study how they evolve over time for the isolated bar models as well the TNG50 galaxies considered here. Sect.~\ref{sec:quadrupole_strength} provides the details of the temporal evolution of the strength and length of the quadrupole features as well as correlation with the corresponding properties of the bar while sect.~\ref{sec:quadrupole_orientation} provides the details of the orientation of the quadrupole feature and their linkage with the bar orientation. 

\subsection{Correlation between strength and length of the bar and the quadrupole feature}
\label{sec:quadrupole_strength}

\begin{figure*}
\centering
\resizebox{0.9\linewidth}{!}{\includegraphics{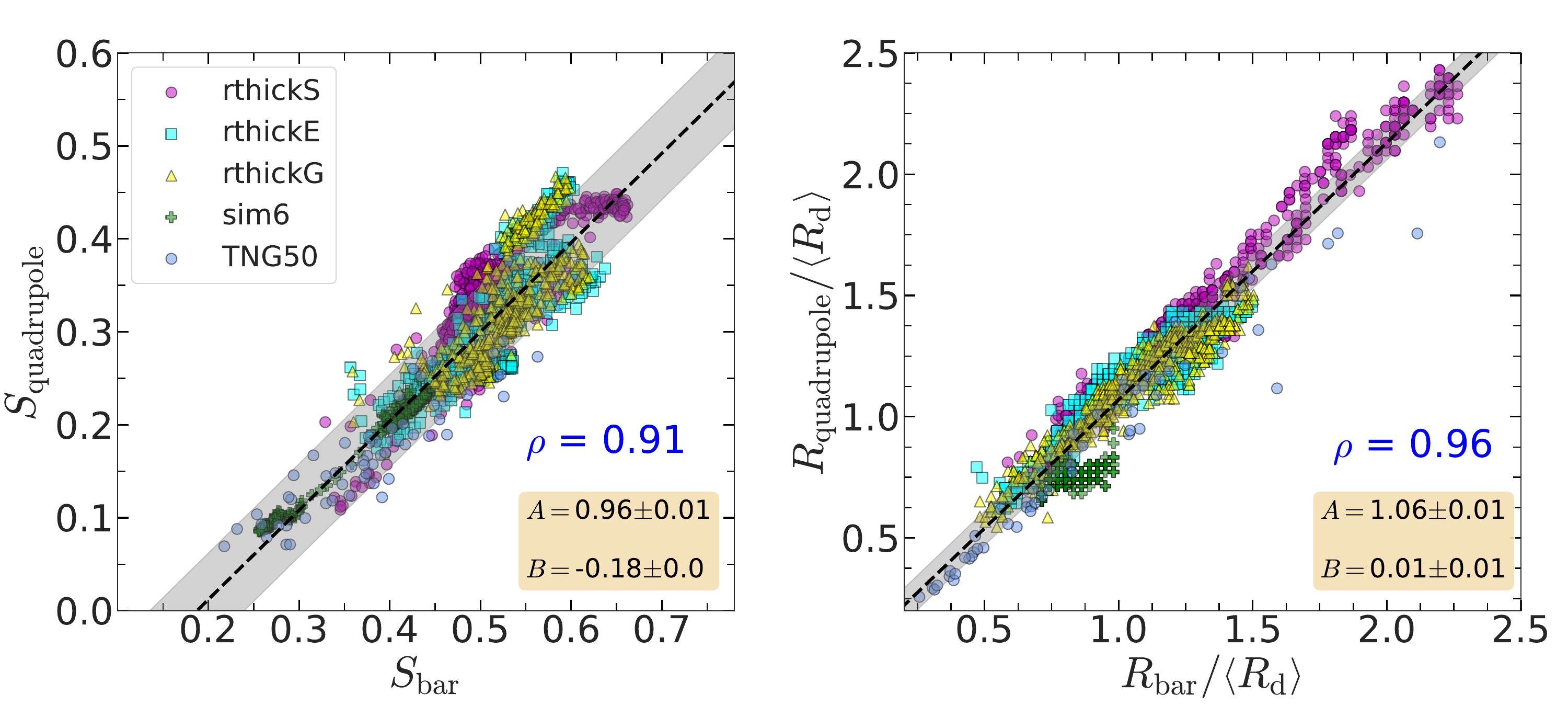}}
\caption{Tracing the bar properties with the quadrupole feature: correlation between the bar strength, $S_{\rm bar}$ and the strength of the quadrupole, $S_{\rm quadrupole}$ (\textit{left panel}), and correlation between the bar length, $R_{\rm bar}$ and the length of the quadrupole, $R_{\rm quadrupole}$ (\textit{right panel}), computed using all isolated thin+thick models and the TNG50 barred galaxies (see the legend). The black dash line denotes the best-fit straight line (of the form $Y = AX+B$) while the grey shaded region denotes the 5-$\sigma$ scatter around the best-fit line. The properties of the bar (strength and extent) remain strongly correlated with the properties of the quadrupole structure (Pearson correlation coefficient, $\rho > 0.75$).}
\label{fig:bar_quadrupole_correlation_allmodels}
\end{figure*}

\begin{figure*}
\centering
\resizebox{0.85\linewidth}{!}{\includegraphics{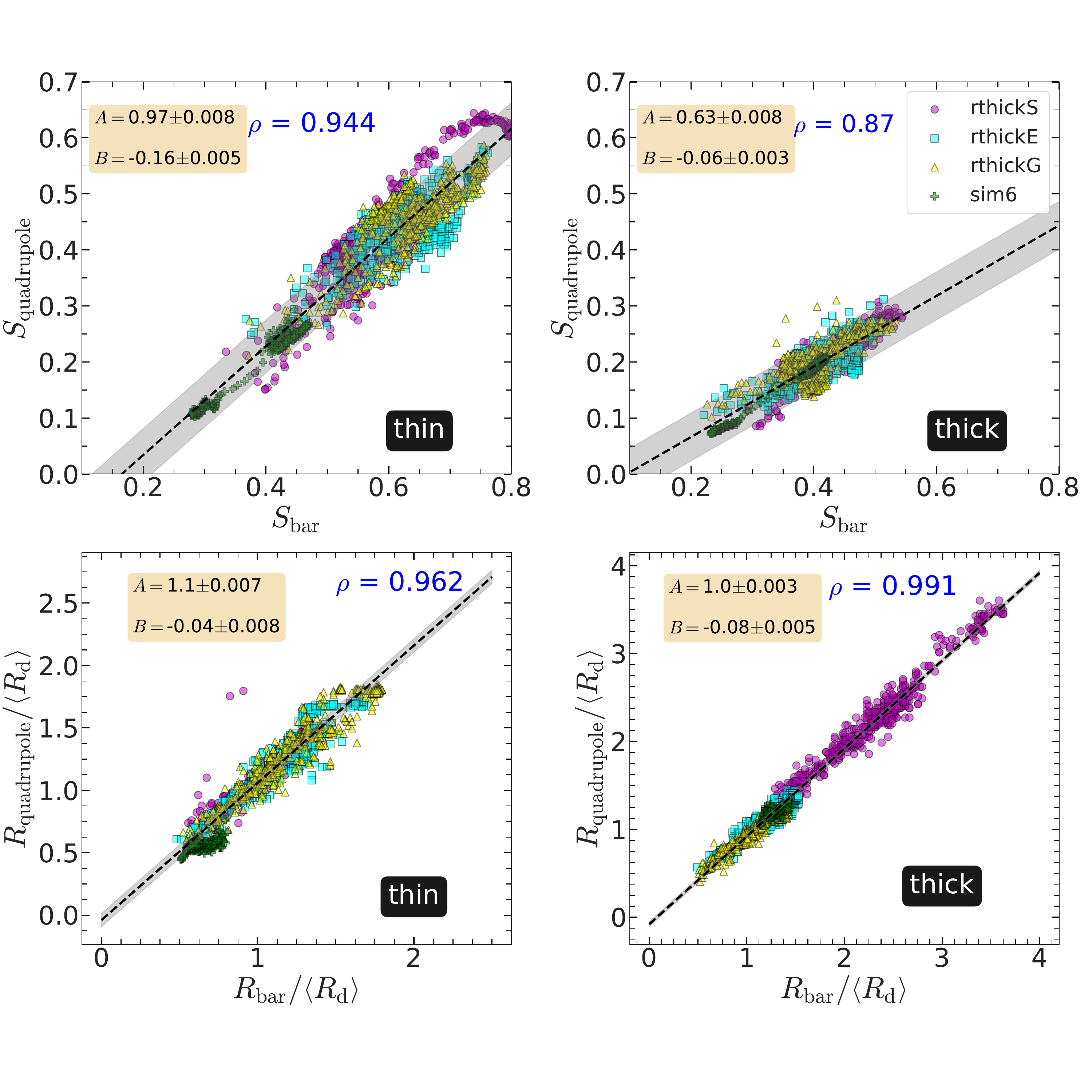}}
\caption{Dependence on the choice of the thin or thick disc stars: correlation between the bar strength, $S_{\rm bar}$ and the strength of the quadrupole, $S_{\rm quadrupole}$ (top row), and correlation between the bar length, $R_{\rm bar}$ and the length of the quadrupole, $R_{\rm quadrupole}$ (bottom row), computed using thin disc particles (left panels) and thick disc particles (right panels), for all thin+thick models and the sim6 model (see the legend). The black dash line denotes the best-fit straight line (of the form $Y = AX+B$) while the grey shaded region denotes the 5-$\sigma$ scatter around the best-fit line. Regardless of the choice of thin or thick disc stellar particles, the length and strength of the quadrupole remain strongly correlated with the length and the strength of the bar.}
\label{fig:bar_quadrupole_correlation_Tracer_dependence}
\end{figure*}


The strength and the length of the bar in our models are computed from the  $m=2$ Fourier coefficients of the underlying density distribution using 
\begin{equation}
A_2/A_0 (R)= \frac{\sum_j m_j e^{i2\phi_j}}{\sum_j m_j}\,.
\label{eq:fourier_calc}
\end{equation}
\noindent Here, $A_2$ denotes the coefficient of the $m=2$ Fourier moment of the density distribution, $m_j$ is the mass of the $jth$ particle, and $\phi_j$ is its azimuthal angle. The summation runs over all the particles within the radial  annulus $[R, R+\Delta R]$, with $\Delta R = 0.5 \kpc$. For each timestep (or equivalently, each snapshot), we calculate the radial profiles of the $m=2$ Fourier coefficient from $R =0$ to $R = 6 R_{\rm d, thin}$. At time $t$, we define the strength of the bar, $S_{\rm bar}$, as the peak value of the $m=2$ Fourier coefficient ($A_2/A_0$). In addition, at time $t$, we define the bar length, $R_{\rm bar}$ as the radial location where $A_2/A_0$ drops to the 70 percent of its peak value in the central bar region. For a detailed exposition to different methods of measuring the length of a bar, the reader is referred to a recent study by \citet{GhoshandDiMatteo2024}. The values of $S_{\rm bar}$ and $R_{\rm bar}$, at different times, for all the thin+thick models are taken from \citet{Ghoshetal2023a} while for the sim6 model and the sample of TNG50 galaxies, we computed the corresponding values of $S_{\rm bar}$ and $R_{\rm bar}$ using the method described above.
\par
The strength and the length of the quadrupole feature in our models are computed from the  $m=4$ Fourier coefficients of the underlying mean radial velocity ($\avg{V_R}$) field using

\begin{equation}
A_4/A_0 (\avg{V_R})= \frac{\sum_j m_j |V_{R,j}| e^{i4\phi_j}}{\sum_j m_j |V_{R,j}|}\,,
\label{eq:fourier_calc_quarupole}
\end{equation}
\noindent where $|V_{R,j}|$ denotes the absolute value of the mean radial velocity of the $jth$ particle. We used the same radial binning, as used in Eq.~\ref{eq:fourier_calc}. Furthermore, the quantity $A_4/A_0 (\avg{V_R})$ is calculated  from $R =0$ to $R = 6 R_{\rm d, thin}$. The corresponding phase angle, $\varphi_4$ is obtained from Eq.\ref{eq:fourier_calc_quarupole} using $\varphi_4 = 1/4 \tan^{-1} (B_4/A_4)$, where $A_4$ and $B_4$ denotes the real and imaginary parts of the summation in Eq.~\ref{eq:fourier_calc_quarupole}. At time $t$, we define the strength of the quadrupole, $S_{\rm quadrupole}$ as the peak value of the $m=4$ Fourier coefficient of the underlying stellar mean radial velocity field ($A_4/A_0 (\avg{V_R})$). The corresponding temporal evolution of the $S_{\rm quadrupole}$ for all thin+thick models as well as for the sim6 model is shown in Appendix~\ref{appen:density_and_Vr_maps} (see top panels of Fig.~\ref{fig:Strength_quadrupole_temporal_allmodels_endstep} there). In addition, at time $t$, we define the length of the  quadrupole, $R_{\rm quadrupole}$, as the radial location where $A_4/A_0 (\avg{V_R})$ drops to the 70 percent of its peak value in the central bar region. Using the same definition, we also computed the corresponding strength and length of the quadrupole feature for the sample of TNG50 barred galaxies considered here.
Furthermore, we recognise that different isolated thin+thick models and the TNG50 barred galaxies used here, have different disc scale lengths. Therefore, for carrying out a uniform comparison, we need to normalise the values of $R_{\rm quadrupole}$ by the corresponding disc scale length. The same argument applies to the bar length ($R_{\rm bar}$) as well. For the thin+thick models and the sim6 model, we measure the (average) disc scale length $\avg{R_{\rm d}}$ using \citep{Ghoshetal2023a}
\begin{equation}
\avg{R_{\rm d}}  = \frac{M_{\rm d, thin} R_{\rm d, thin} + M_{\rm d, thick} R_{\rm d, thick}} {M_{\rm d, thin}+M_{\rm d, thick}}\,,
\label{eq:avg_scalelength}
\end{equation}
\noindent where $M_{d, j}$ denotes the stellar mass and $R_{d, j}$ denotes the scale length of the $jth$ component ($j = $ thin, thick). As for the TNG50 barred galaxies, we fit a single exponential profile of the form $\Sigma (R) \propto \Sigma_0 \exp[- R/R_{\rm d}]$ to the surface density profiles along the bar major axis while excluding the central bar region \footnote{For a few TNG50 galaxies, the surface density profiles show a break in the outer disc region, therefore a more rigorous approach would have required to fit a double-exponential profile. However, for this work, we used only a single exponential profile.}.
\par
Next, we investigate if there exists any correlation between the strength and length of the quadrupole feature and the bar. Fig.~\ref{fig:bar_quadrupole_correlation_allmodels} (left panel) shows the corresponding correlation between the bar strength, $S_{\rm bar}$ and the quadrupole strength $S_{\rm quadrupole}$ for all isolated models and the TNG50 galaxies considered here. We mention that in different thin+thick models, bar forms at different times \citep[for further details, see][]{Ghoshetal2023a,Ghoshetal2024}. However, we checked that a prominent bar is always associated with a quadrupole feature, regardless of its formation time. Therefore, only the snapshots after the bar forms are considered for all the thin+thick as well as for the sim6 model. In addition, to quantify whether the quantities $S_{\rm bar}$ and $S_{\rm quadrupole}$ are correlated or not, we compute the corresponding Pearson correlation coefficient, $\rho$, and find that indeed these two quantities are strongly correlated ($\rho \geq 0.75$). In other words, the bar and the quadrupole feature in the stellar velocity field evolve in tandem, and this trend holds for all the models considered here. Furthermore, we fit a straight line of the form $Y = AX+B$ to all the points in the $S_{\rm bar}$-$S_{\rm quadrupole}$ plane to check if these two quantities follow a linear scaling law. The best-fit parameters ($A = 0.96 \pm 0.1$; $B = -0.18 \pm 0.001$) suggest that the strength of quadrupole are indeed linearly related to the bar strength.  Fig.~\ref{fig:bar_quadrupole_correlation_allmodels} (right panel) shows the corresponding correlation between the bar length, $R_{\rm bar}$ and the quadrupole length, $R_{\rm quadrupole}$ for all isolated models and the TNG50 galaxies considered here. The calculated Pearson correlation coefficient, $\rho$ is found to be greater than 0.75; thereby demonstrating that these two quantities are also strongly correlated. We fit a straight line (of the form $Y = AX+B$) to all the points in the $R_{\rm bar}$-$R_{\rm quadrupole}$ plane (both quantities being normalised by the same average disc scale length, $\avg{R_{\rm d}}$). As the best-fit parameters ($A = 1.06 \pm 0.1$; $B = 0.01 \pm 0.01$) suggest, the lengths of the bar and the quadrupole remain linearly related to each other.
\par
Lastly, we investigate whether the choice of kinematically-cold thin disc or kinematically-hot thick disc stars can affect the earlier found correlations between the properties of the bar and the quadrupole feature. We mention that, in the sim6 and all the thin+thick models, we can identify and separate, by construction, which stars are members of the thin disc component at initial time ($t=0$) and which stars are members of the thick disc component at $t=0$, and we can track them as the system evolves self-consistently. This, in turn, allows us to test the dependence (if any) of the correlations found between the properties of the bar and the quadrupole feature, on the choice of thin disc or thick disc stars. To achieve that, we recalculate the strength and the length of both the bar and the quadrupole feature, using the thin and thick disc particles separately. The corresponding correlations, as a function of stellar populations with varied degree of velocity dispersion (i.e. kinematically-colder thin disc and kinematically-hotter thick disc) \footnote{for the sim6 model, both intermediate and thick disc stars are together considered as `thick' disc population.} are shown in Fig.~\ref{fig:bar_quadrupole_correlation_Tracer_dependence}. As seen clearly from Fig.~\ref{fig:bar_quadrupole_correlation_Tracer_dependence}, both the length and the strength of bar and the quadrupole feature remain strongly correlated (Pearson correlation coefficient, $\rho > 0.75$), regardless of the choice of thin disc or thick disc stars. As for the linear scaling relation, the choice of thin disc or thick disc stars does not change appreciably (less than 10 percent) the value of the best-fit slope of straight line when fitted to $R_{\rm bar}$ and $R_{\rm quadrupole}$ (see bottom panels in Fig.~\ref{fig:bar_quadrupole_correlation_Tracer_dependence}). However, for the quantities  $S_{\rm bar}$ and $S_{\rm quadrupole}$, the best-fit straight line is shallower when computed using only the thick disc stars, as compared to when they were calculated using only the thin disc stars (see the corresponding best-fit values in top panels of Fig.~\ref{fig:bar_quadrupole_correlation_Tracer_dependence}). As shown in \citet{Ghoshetal2023a}, the thin disc stars constitute a stronger bar and the thin disc stars (being the kinematically colder component) are more perturbed kinematically by the non-axisymmetric structures (e.g. $m=2$ bar here, also see \citet{Debattistaetal2017}). Therefore, a steeper best-fit straight line for the thin disc stars in the $S_{\rm bar}$-$S_{\rm quadrupole}$ plane is likely to be a combined result of the two above-mentioned dynamical effects.

\subsection{Orientation of quadrupole and the bar}
\label{sec:quadrupole_orientation}

\begin{figure*}
\centering
\resizebox{0.95\linewidth}{!}{\includegraphics{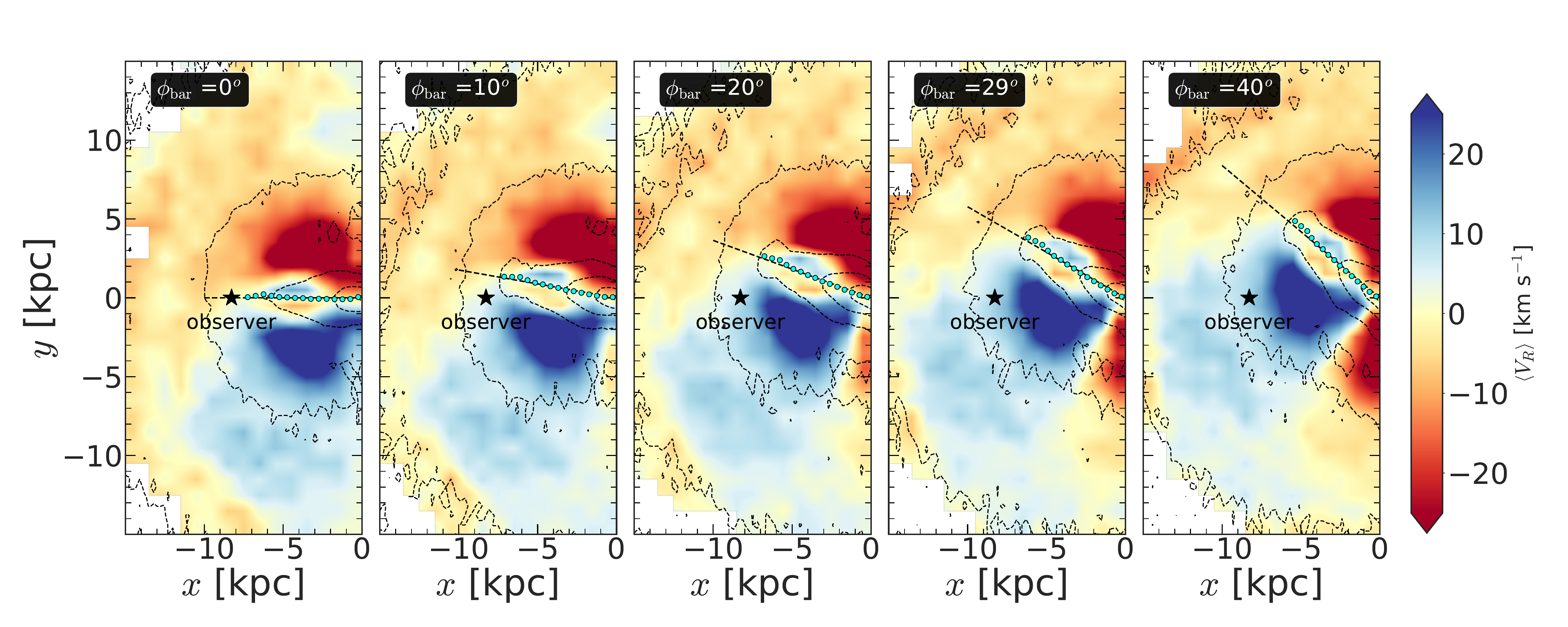}}
\caption{Distribution of the mean radial velocity ($\avg{V_R}$) in the $(x-y)$-plane, calculated at $t=9 \Gyr$ for the model rthick0.0, with the bar placed at different viewing angles with respect to a hypothetical observer (shown in diamond) at a Solar-like position ($R = -8 \kpc$, $\phi = 0 \degrees$, $z = 0$). The cyan circles in each sub-panel denote the variation of the phase-angle of the $m=4$ Fourier coefficient ($\varphi_4$), computed from the distribution of $\avg{V_R}$. The dashed straight line denotes the `true' bar angle, computed from the intrinsic density distribution of the stellar particles.}
\label{fig:bar_quadrupole_orientation_example}
\end{figure*}

\begin{figure}
\centering
\resizebox{0.825\linewidth}{!}{\includegraphics{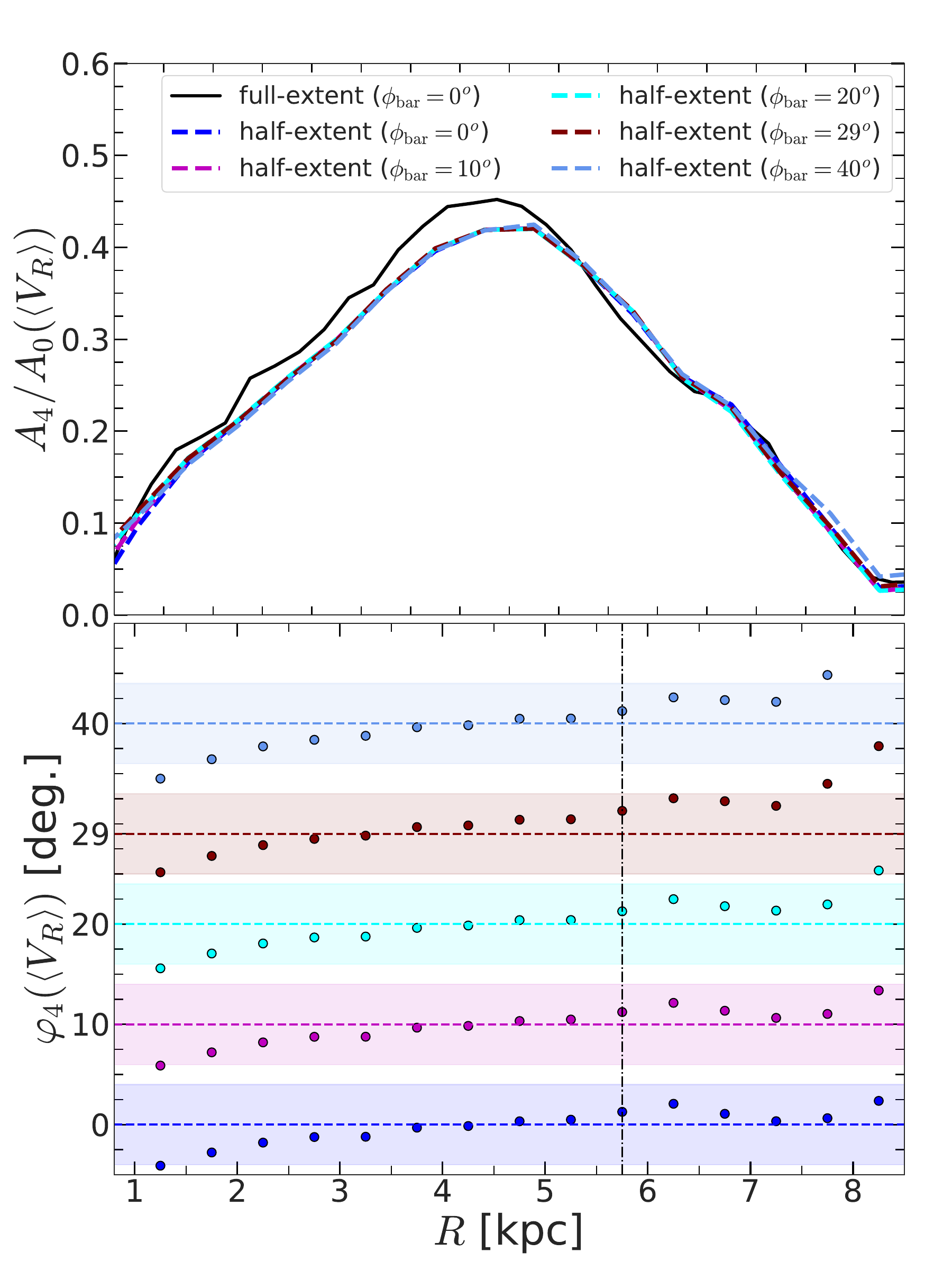}}
\caption{Top panel: Radial variation of the amplitude of the $m=4$ Fourier moment of the mean radial velocity, $\avg{V_R}$, calculated at $t = 9 \Gyr$ for the model rthick0.0, while putting the bar at different orientations (see the legend). Bottom panel: radial variation of the corresponding phase-angle ($\varphi_4$) of $m=4$ Fourier moment of the mean radial velocity, $\avg{V_R}$. The points denoting the radial variation computed from the mean radial velocity, $\avg{V_R}$ while the horizontal lines denote the corresponding true bar orientation, and the shaded region around the true value denotes a $8\degrees$ scatter. `Half-extent' refers to the scenario where stars falling only in the negative half ($x \leq 0$) are considered while computing the Fourier moments. For further details, see the text. The vertical dash-dotted black line denotes the length of the bar ($R_{\rm bar} = 5.75 \kpc$).}
\label{fig:bar_quadrupole_orientation_quantification}
\end{figure}

In the earlier section, we demonstrated that the strength and the extent of the quadrupole feature in the $\avg{V_R}$ distribution is an excellent proxy for the strength and the length of a bar. We mention that the Fourier decomposition of the mean radial velocity distribution (see Eq.~\ref{eq:fourier_calc_quarupole}) not only provides the amplitude of the $m=4$ quadrupole feature, but the corresponding phase angle ($\varphi_4$) also provides the information about the orientation of the quadrupole feature. 
In Fig.~\ref{fig:density_maps_endstep_allmodels}, we show the spatial variation of the phase angle of the $m=4$ Fourier moment ($\varphi_4$) of the $\avg{V_R}$ distribution, within and beyond the bar region. Even a mere visual inspection reveals that the phase angle, $\varphi_4$, remains constant within the bar region, and the distribution of the $\varphi_4$ follows the orientation of the bar. We checked that this trends hold true for all models considered here; thereby implying that the orientation of the quadrupole feature in the $\avg{V_R}$ can be used as a proxy for the bar orientation in the density field. However, in the MW, we almost do not see the other side of the Galactic centre; thereby making it difficult to reconstruct the full quadrupole feature in the corresponding mean radial velocity field. The question remains: in this case (and for arbitrary orientations of the bar), how robustly the orientation of the quadrupole feature (in $\avg{V_R}$ distribution) can trace or recover the bar orientation (in the stellar density field). Here, we test this in detail.
\par%
Fig.~\ref{fig:bar_quadrupole_orientation_example} shows an example of how the distribution of $\avg{V_R}$ in the $(x-y)$-plane, would appear to a hypothetical observer at a Solar-like position ($R = -8 \kpc$, $\phi = 0^o$, $z = 0$) when the bar is placed at different orientations. In addition, to mimick a MW-like situation, we have only considered the negative-half ($x \leq 0$) of the distribution of $\avg{V_R}$ in the $(x-y)$-plane. We then repeat the Fourier decomposition of the distribution of $\avg{V_R}$ (taking only the negative half and bar placed at different orientations) to recompute the strength and the extent of the quadrupole. Fig.~\ref{fig:bar_quadrupole_orientation_quantification} shows the corresponding strength and orientation of the quadrupole while placing the bar at different angles. As seen from Fig.~\ref{fig:bar_quadrupole_orientation_quantification}, the quadrupole strength is recovered within 10 percent (relative) errors for all assumed bar orientations and considering the negative-half ($x \leq 0$). The quadrupole orientation ($\varphi_4$) also recovers the bar orientations, albeit with lesser accuracy. Further, the $\varphi_4$ values in the central regions, show a departure from the `true' values (however, they remain within $\pm 4 \degrees$). We noticed that, for larger bar orientation angles ($\phi_{\rm bar} > 10 \degrees$), the $\avg{V_R}$ distribution no longer remains bi-symmetric (wrt. $x=0$ line) as the part of the lobe from the other-half (i.e. $x >0$) starts appearing to the hypothetical observer at a Solar-like position (compare cases for $\phi_{\rm bar}= 0 \degrees$ and  $\phi_{\rm bar}= 40 \degrees$ in Fig.~\ref{fig:bar_quadrupole_orientation_example}).  This, in turn, introduces a systematic fluctuation (around the true value of $\phi_{\rm bar}$) in inferring the bar orientation angle from the measured $\varphi_4$ values (see the bottom panel of Fig.~\ref{fig:bar_quadrupole_orientation_quantification}).

\subsection{Influence of spiral arms in measuring the quadrupole length}
\label{sec:bar_spiral_conundrum}

\begin{figure*}
\centering
\resizebox{1.025\linewidth}{!}{\includegraphics{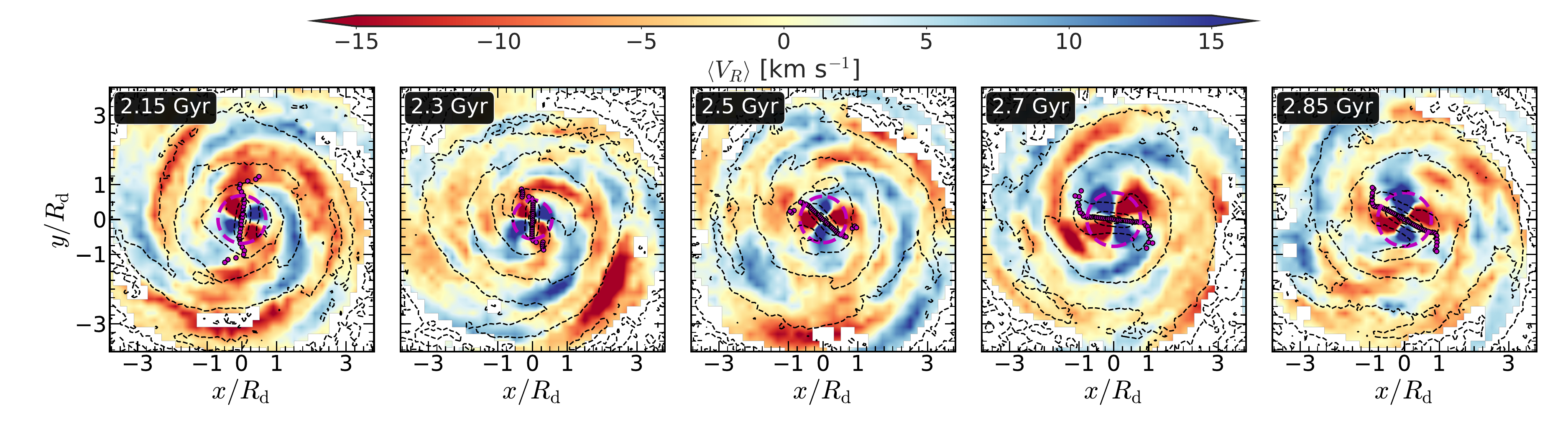}}
\caption{Distribution of the mean radial velocity, $\avg{V_R}$ in the $(x-y)$-plane, computed at different times when the model rthick0.0 harbours a strong bar and prominent spirals.  Black dashed lines denote the contours of constant surface density. The magenta points denote the spatial distribution of the phase-angle of the $m=4$ Fourier moment ($\varphi_4$). Within the bar region, $\varphi_4$ remains constant whereas in presence of prominent spirals, $\varphi_4$ shows a characteristic modulation, and this trend holds for all bar+spiral scenarios shown here.}
\label{fig:bar_spiral_conundrum_example}
\end{figure*}


\begin{figure}
\centering
\resizebox{\linewidth}{!}{\includegraphics{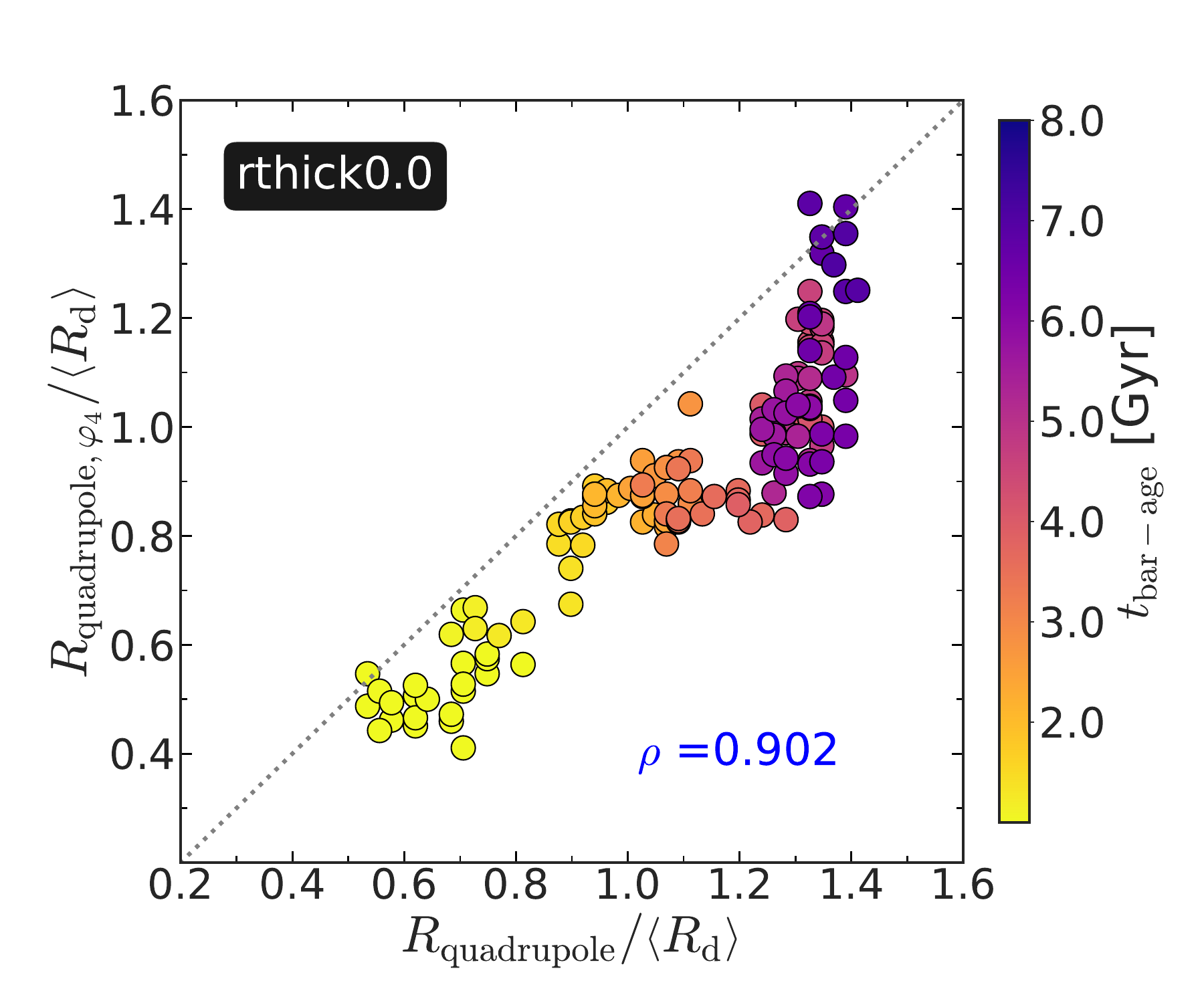}}
\caption{Correlation between the two methods of measuring the quadrupole extent : $R_{\rm quadrupole}$, measured solely from the amplitude of the $m=4$ Fourier coefficient of the mean radial velocity $\avg{V_R}$, and $R_{\rm quadrupole, \varphi_4}$, measured solely from the constancy of the phase-angle, $\varphi_4$ of the mean radial velocity $\avg{V_R}$, as a function of bar age ($t_{\rm bar-age}$, see the colour bar). Both the quantities, $R_{\rm quadrupole}$ and $R_{\rm quadrupole, \varphi_4}$, are normalised by the mean disc scale length, $\avg{R_{\rm d}}$. For details, see the text. The dotted straight line denotes the 1:1 relation. The colour bar denotes the age of the bar. }
\label{fig:bar_spiral_conundrum_length_variation}
\end{figure}


In previous sections, we demonstrated that the properties of the quadrupole feature are well correlated with the properties of bar. However, the MW harbours other non-axisymmetric features, for example, spirals \citep[e.g.,][]{Oort1958,GeorgelinandGeorgelin1976,Gerhard2002,Churchweletal2009,Reidetal2014} which also excite non-zero mean radial velocities in the disc region \citep[e.g. see][]{Siebertetal2011,Siebertetal2012}. Interestingly, the non-zero mean radial velocity, excited by spirals, can often overlap or is connected spatially with the non-zero mean radial velocity excited by the bar (i.e. the quadrupole feature) \citep[e.g. see][]{Visloskyetal2024}. This, in turn, can pose a problem in disentangling the dynamical effect of the bar on mean radial velocities, and can result in overestimating the extent of the quadrupole feature. This is similar to the dynamical situation where spirals emerge from the end tip of the bar, causing an overestimation of the bar length \citep[for a detailed discussion, see][]{Hilmietal2020,GhoshandDiMatteo2024}. In such a dynamical situation, using the constancy of the phase-angle ($\varphi_2$) of the $m=2$ Fourier moment (of the density distribution) can potentially decrease the overestimation of the bar length due to the presence of spirals, as demonstrated in \citet{GhoshandDiMatteo2024}. Here, we pursue a similar strategy. 
\par
Fig.~\ref{fig:bar_spiral_conundrum_example} shows the distribution of $\avg{V_R}$ in the $(x-y)$-plane at different times for the model rthick0.0 which harbours a bar+spiral feature. As seen clearly from a visual inspection, non-zero $\avg{V_R}$ in the disc region dominated by the spirals, are connected to the quadrupole feature excited by the bar. Fig.~\ref{fig:bar_spiral_conundrum_example} further shows the distribution the $m=4$ Fourier phase-angle (from the $\avg{V_R}$ distribution) in the disc regime where the $m=2$ bar is dominant as well as in the outer disc region where the spirals are dominant. In all such bar+spiral scenarios, the the phase-angle, $\varphi_4$ remains constant (within $5-8 \degrees$) in the central bar region (also see Fig.~\ref{fig:density_maps_endstep_allmodels}) whereas in the outer disc region (dominated by the spirals), the phase-angle, $\varphi_4$ does not remain constant. In other words, phase-angle, $\varphi_4$ displays a characteristic modulation as one moves out from the central bar dominated region to outer spiral dominated region. \citet{GhoshandDiMatteo2024} showed a similar characteristic change in the $m=2$ phase-angle (from the density distribution) for a bar+spiral asymmetry. Therefore, we adopt a new (conservative) definition of the extent of the quadrupole, $R_{{\rm quadrupole}, \varphi_4}$ as the radial extent within which the $m=4$ Fourier phase-angle remains constant (within $\sim 5-8 \degrees$). Fig.~\ref{fig:bar_spiral_conundrum_length_variation} shows the corresponding comparison between the two different length estimators for the quadrupole feature, namely, $R_{{\rm quadrupole}}$ and $R_{{\rm quadrupole}, \varphi_4}$, at different times for the model rthick0.0. As seen from Fig.~\ref{fig:bar_spiral_conundrum_length_variation}, the values of $R_{{\rm quadrupole}, \varphi_4}$ remains systematically lower than the values of $R_{{\rm quadrupole}}$ when the model harbours both a stellar bar and (transient) spirals. Only towards the end phase of the evolution, when the model no longer hosts spirals, these two values match fairly well (see Fig.~\ref{fig:bar_spiral_conundrum_length_variation}). In other words, the presence of (transient) spirals systematically overestimates the length of the quadrupole feature (with a median relative difference of $\sim 25$ percent).  We checked that this trend holds true for other models as well which display a bar+spirals scenario. For the sake of brevity, they are not shown here.  Therefore, the results presented here, outlines the importance of using $R_{{\rm quadrupole}, \varphi_4}$ as a more appropriate estimator for measuring the length of the quadrupole (and, in turn, inferring the bar length) in a bar+spirals dynamical scenario.

\section{Implication on constraining the MW's bar properties from the quadrupole feature}
\label{sec:MW_bar_properties}

In previous sections, we have demonstrated that for simulated bars (from a wide variety of numerical simulations), the quadrupole feature in the stellar mean radial velocity distribution serves as an excellent kinematic diagnostic for constraining the bar's properties. However, it remains to be tested whether the quadrupole feature still works as a kinematic diagnostic in presence of dust extinction and \gaia\ -like uncertainties and systematic errors for the MW. We pursue it here. In sect.~\ref{sec:MW_dataset}, we describe the sample selection from the \gaia\ DR3 whereas in sect.~\ref{sec:gaia_like_testing}, we provide the details of creating a \gaia\ -like mock dataset from the high-resolution sim6 model (for details, see sec.~\ref{sec:isolated_abrredmodel}). Lastly, the findings are provided in sect.~\ref{sec:application_to_MW}.

\subsection{Sample selection from Gaia DR3}
\label{sec:MW_dataset}

\begin{figure*}
    \centering
    \includegraphics[width=0.85\linewidth]{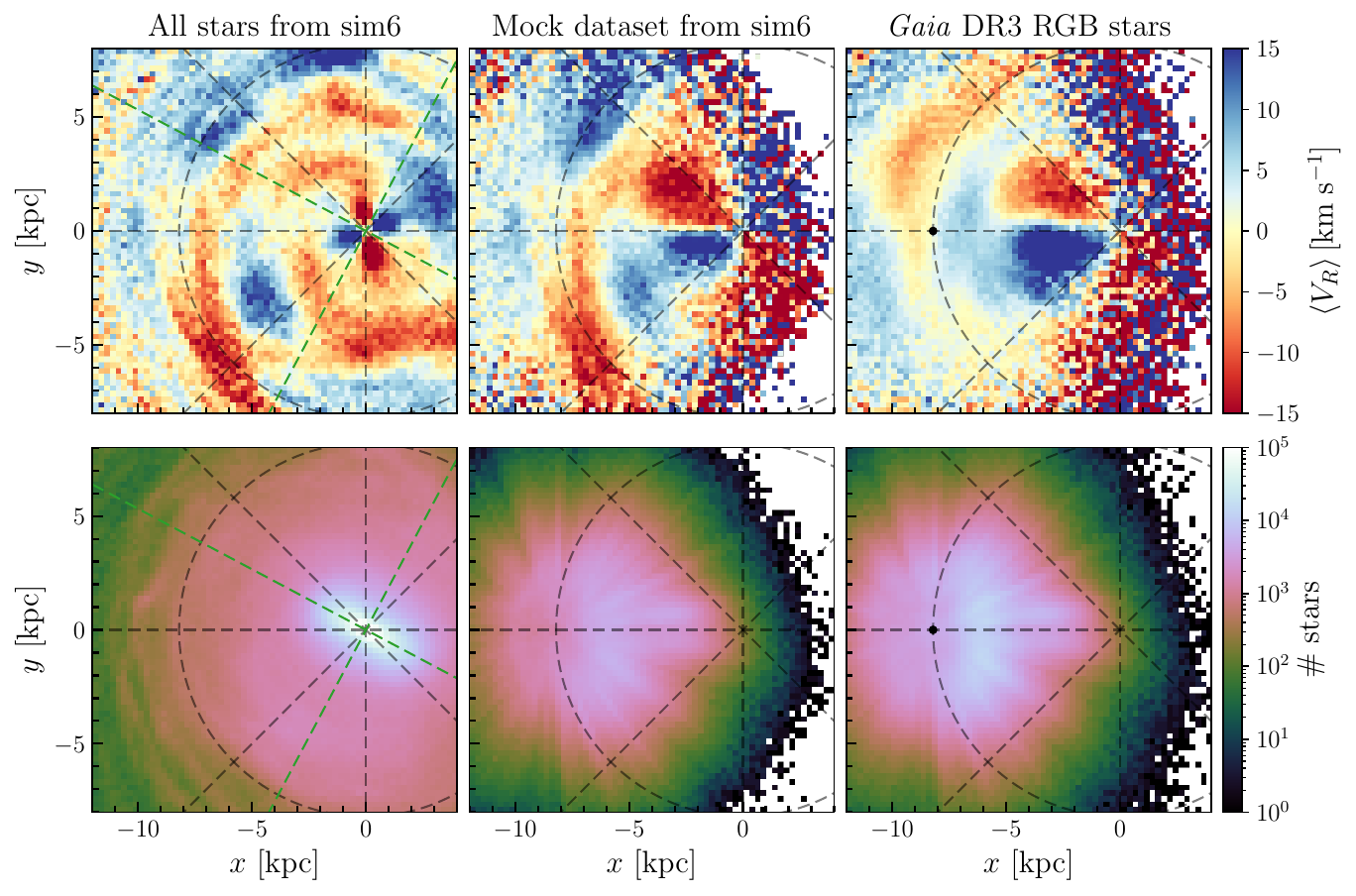}
    \caption{Comparison between the \gaia-like mock data and the \gaia\ DR3 sample : Top panels showing the $\avg{V_R}$ distribution and the bottom panels showing the stellar density, both in the $(x-y)$-plane, calculated for all the stellar particles from model sim6 (left column), the \gaia-like mock dataset from model sim6 (middle column) and the selected sample of RGB stars from the \gaia\ Data Release 3. As seen evidently, the mock dataset qualitatively reproduces the behaviour seen in the \gaia\ sample, notably, observational effects increase the extent of the quadrupole and reduce the angle of its main axis compared to $y=0$. The green lines in the left column corresponds to the major and minor axis of the bar, angled at $28^\circ$. The black filled circle in the right column denotes the assumed Solar location (for further details, see sect.~\ref{sec:MW_dataset}).}
    \label{fig:sim0300_vs_RGB_comparison}
\end{figure*}

We select a sample of RGB stars from the \gaia\ DR3, following broadly the same procedure as described in \cite{Drimmeletal2023}, by selecting stars with $3000\,\mathrm{K} < T_\mathrm{eff} < 5500\,\mathrm{K}$ and $\log g < 3.0$. We limit ourselves to stars with full 6-D kinematics and apparent magnitudes within the brightness limits $5 < G < 15.5$ that have high fidelity astrometric solutions ($\texttt{fidelity\_v2}>0.5$ from \cite{Rybizki21}). The parallax zero points are corrected using the method described in \cite{Groenewegen21}, and the sample is augmented with extinction $E$ using the all-sky extinction map from \cite{Zhang24}. The extinction map extends out to $5\mathrm{\ kpc}$ (from the Solar location), beyond which it is extrapolated using a double-exponential model for the differential dust density. We further apply quality cuts $\texttt{radial\_velocity\_error} < 20\ \kms$ and $\varpi/\sigma_\varpi > 5$. We calculate distance using the simple inverse of the measured parallax, $d = 1/\varpi$. The positions and velocities of the stars are transformed to Galactocentric coordinates using the \texttt{astropy.coordinates} module \citep{astropy22}. For the transformation, we use $z_{\odot}=20.8\pc$ \citep{Bennett19}, $R_{\odot}=8.277 \kpc$, and $v_\odot=(9.3\kms,251.5\kms,8.59\kms)$ \citep{Gravity22}. The Sun is then located at $(-R_{\odot}, 0, z_{\odot})$ moving with velocity $v_\odot$. Our final selected sample contains 5,667,443 stars. The corresponding stellar density and the $\avg{V_R}$ distribution in the Galactocentric $(x-y)$-plane, computed for our final selected sample from the \gaia\ DR3 are shown in  Fig.~\ref{fig:sim0300_vs_RGB_comparison} (see the right column). A prominent quadrupole feature is present in the $\avg{V_R}$ distribution calculated in the $(x-y)$-plane, in agreement with  \cite{Drimmeletal2023}.

\subsection{Building Gaia-like mock datasets}
\label{sec:gaia_like_testing}

Next, in order to investigate any plausible effect of the \gaia\ -like uncertainties and systematic errors on the robustness of the quadrupole feature as a kinematic diagnostic, we first generate {\it Gaia}-like mock datasets using the snapshots from the sim6 model at different times (and thus for different bar strengths; see Fig.~\ref{fig:Strength_quadrupole_temporal_allmodels_endstep}). For each snapshot, we place a hypothetical observer at a Solar-like position at $(-R_{\odot}, 0, z_{\odot})$, and at a bar viewing angle of $\phi_{\rm bar} = 28\degrees$ that imitates the qualitative behaviour of a sample of RGB stars in \gaia\ DR3.

Next, we model the errors in measured parallax, radial velocity, proper motion, and apparent magnitude. The observational errors for each simulated star are taken from randomly chosen stars in the \gaia\ DR3 sample with apparent magnitude $G > 15$. The apparent magnitude threshold is chosen because most of the stars in \gaia\ DR3 for which the effect of observational errors is significant are further away and have their apparent magnitudes in the corresponding range. Observed values of parallax, radial velocity, and proper motion are then simulated by adding normally distributed errors to the true values, with standard deviations as determined above: $X_\mathrm{observed}\sim \mathcal{N}(\mu=X_\mathrm{true}, \sigma^2=X_\mathrm{error}^2)$ for an observable $X$. The above process is repeated ten times, in order to end up with a comparable number of particles as is in the {\it Gaia} dataset.

In order to account for the selection function, we adopt the same procedure as in \cite{Zhang_Hanyuan_23}. This is done by matching each real star in the \gaia\ DR3 dataset with its closest counterpart in the simulation. Simulated particles that failed to obtain a match or that are more than 0.1\,kpc away from the matched real star are discarded. The final mock dataset has a spatial distribution that closely mimics that of the \gaia\ DR3 sample (see Fig.~\ref{fig:more_on_GaialikeMock} in Appendix~\ref{appen:more_on_GaialikeMocks}). This approach has the benefit of implicitly accounting for complicated selection effects such as crowding, dust, and the \gaia\ scanning law. However, it can fail to properly account for the origin of the particles. For example, in \gaia, a star hidden behind a thick dust cloud will never be observable due to the dust cloud blocking the light, but in the mock dataset it could radially scatter into the visible volume.The $\avg{V_R}$ distribution for one of the mocks, calculated at $t=3\Gyr$ for the sim6 model, is shown in Fig.~\ref{fig:sim0300_vs_RGB_comparison} (see middle column).

\begin{figure*}
    \centering
    \includegraphics[width=0.9\linewidth]{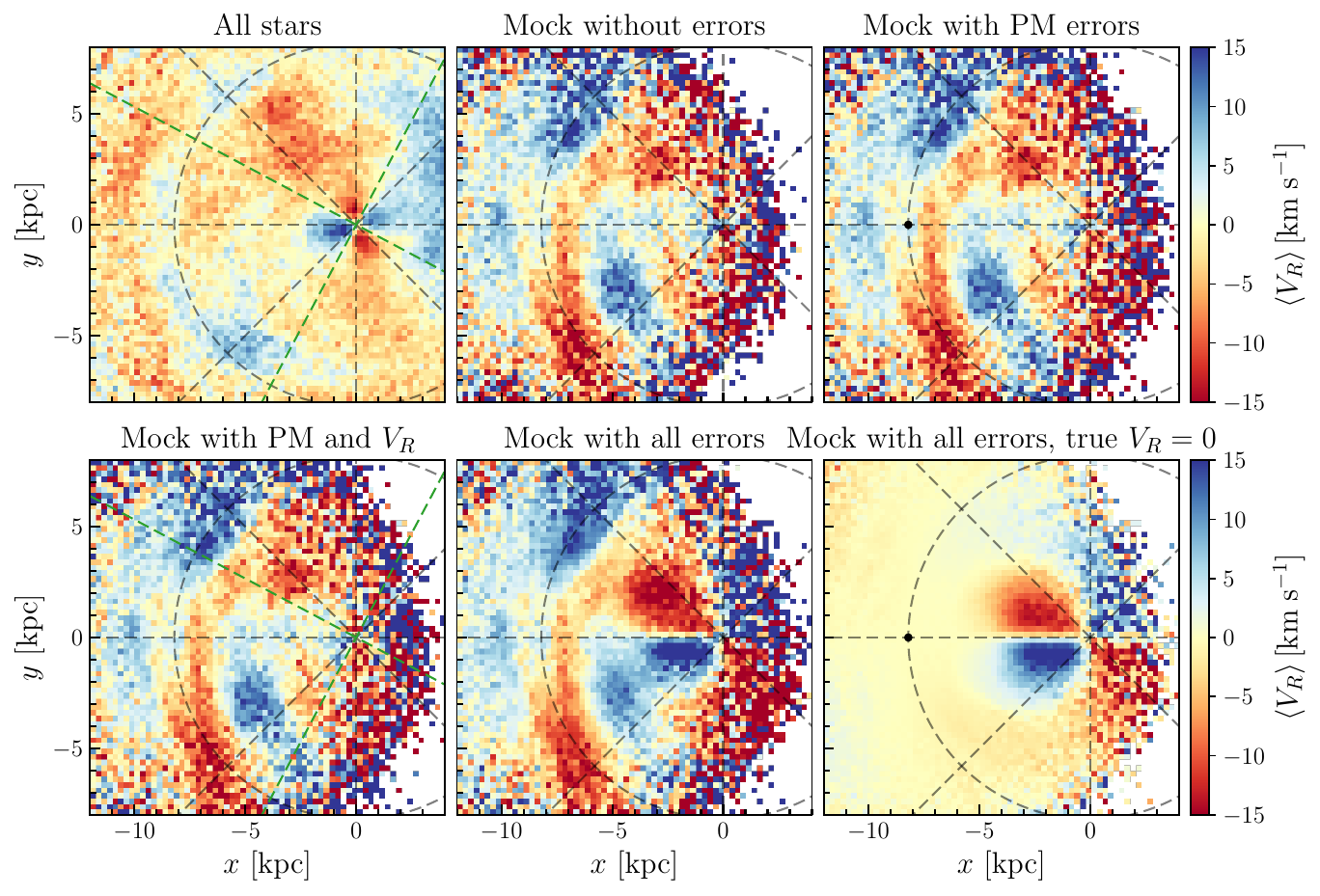}
    \caption{Investigating the effects of different uncertainties: Distribution of $\avg{V_R}$ in the $(x-y)$-plane for the \gaia-like mock dataset at $t=3\mathrm{\,Gyr}$, showcasing how the observed quadrupole feature is affected by errors in proper motion, radial velocity ($V_R$), and parallax. The top left panel is for stars in the snapshot (no uncertainties and no magnitude cut) while top middle panel shows the magnitude limited $G \leq 15.5$ mock dataset with no errors. The top right panel and the bottom left panel show the mock datasets with only proper motion (PM) uncertainty and both the PM and radial velocity uncertainties, respectively. The bottom middle panel shows the mock dataset resampled according to the errors in the \gaia\ DR3 RGB subsample (see section \ref{sec:gaia_like_testing}), and the bottom right panel shows the mock dataset where all the particles have had their $V_R$ set to zero prior to applying observational effects. For further details, see sect.~\ref{sec:application_to_MW}.}
    \label{fig:0300_parallax_contamination_analysis}
\end{figure*}

\subsection{Quadrupoles in  Gaia-like mocks: results and limitations}
\label{sec:application_to_MW}

First, we choose the snapshot at $t = 3 \Gyr$ from the sim6 model to investigate the properties of the quadrupole in $\avg{V_R}$ in \gaia-like mock datasets. We mention that there is nothing special about the choice of $t= 3 \Gyr$ except the fact that it harbours a prominent stellar bar in the central region. Later, we have tested the same robustness of quadrupole in $\avg{V_R}$ for other snapshots as well.
Looking at the distribution of $\avg{V_R}$ in the $(x-y)$-plane for the MW RGB subsample and the mock dataset highlighted in Fig. \ref{fig:sim0300_vs_RGB_comparison}, we observe largely the same behaviour of a prominent and extended quadrupole signature near the galactic center which tapers off around $R=4\kpc$. For the simulated model, even a visual inspection reveals that the bar angle in the mock data is lying closer to the $y=0$ axis (as derived from the orientation of the quadrupole as $\varphi_4 \sim 5 \degrees$), in sharp contrast with the true bar angle of $\sim 28 \degrees$. In addition, using the methods described in sect.\ref{sec:quadrupole_strength}, we calculate the strength of the quadrupole from the mock dataset to be $S_\mathrm{quadrupole} = 0.37$ and the length $R_\mathrm{quadrupole}=3.8\kpc$. When compared with the `true' values (i.e. directly calculated from the simulation at $t = 3 \Gyr$), we find that  the $S_\mathrm{quadrupole}$ is overestimated by $\sim 40$ percent and the $R_\mathrm{quadrupole}$ is overestimated by $\sim 35$ percent. We repeated this procedure for a few other snapshots which harbour prominent bars. We find that the overestimation in inferring $S_\mathrm{quadrupole}$ varies in the range $40-45$ percent (of the `true' value), and for $R_\mathrm{quadrupole}$, the corresponding overestimation varies in the range $35-45$ percent (of the `true' value). Therefore, the values of the quadrupole's properties are not well recovered from the \gaia-like mock data.  The large difference between the values derived from the \gaia-like mock dataset and the true values (directly obtained from the simulation), is quite puzzling and warrants further investigation. Next, we pursue this. 
\par
In Fig.~\ref{fig:0300_parallax_contamination_analysis}, we demonstrate how the distribution of $\avg{V_R}$ in the $(x-y)$-plane, changes in the mock dataset as we sequentially incorporate the uncertainties from proper motion, radial velocity, and parallax measurements. As seen from Fig.~\ref{fig:0300_parallax_contamination_analysis} (top middle panel), when the mock data does not include any uncertainties, but is completeness limited, it still shows a quadrupole feature, albeit fainter. The weakening of the quadrupole is due to the effect of dust (implicit in the applied magnitude cut) which essentially blocks the stars towards the Galactic centre and closer to the mid-plane. Since the effect of bar decreases as one moves away from the mid-plane, therefore the quadrupole structure, imprinted in the kinematics of the stars (far away from the mid-plane), becomes weaker. The introduction of uncertainties in radial velocity and proper motion does not drastically change the overall distribution of $\avg{V_R}$ (see the top right and bottom left panels in Fig.~\ref{fig:0300_parallax_contamination_analysis}).
We find that the bulk of the effect from observational errors comes from the uncertainty in parallax. A signal-to-noise ratio of 10 causes an uncertainty of $\sim 1\kpc$ along the line of sight near the Galactic centre. Since $V_R=\vec V\cdot \hat{R}$ is the projection of a star's velocity on the cylindrical radial unit vector, which has a singularity at the Galactic centre, $V_R$ gets contaminated by other velocity components as results of the shift of its observed position around the Galactic Centre. In Fig.~\ref{fig:0300_parallax_contamination_analysis} (bottom right panel), we further demonstrate how a scenario where the effect of the intrinsic $V_R$ of the particles is eliminated by setting all the particles to artificially have zero $V_R$ prior to the application of observational effects reproduces the central observed quadrupole. Thus, the `observed' quadrupole present in the final mock dataset (bottom middle panel of Fig.~\ref{fig:0300_parallax_contamination_analysis}) is essentially a manifestation of the intrinsic fainter quadrupole which is greatly enhanced by the uncertainties in parallax measurement. This clearly outlines the pitfall of inferring MW's bar properties by using the stellar kinematic information from the \gaia\ DR3 when proper care is not taken into account for the observational biases of the \gaia\ survey.
\par
Lastly, we measure the strength and the length of the quadrupole (for definitions, see sect.~\ref{sec:quadrupole_strength}) for the MW, using our selected sample of RGB stars (for details of the sample selection, see sect.~\ref{sec:MW_dataset}). The strength of the quadrupole $S_\mathrm{quadrupole}=0.37$ and the extent $R_\mathrm{quadrupole}=4.75\kpc$ for the selected sample of the RGB stars in the MW. Now, if one assumes the empirical relations for the $S_{\rm bar} - S_{\rm quadrupole}$ and the $R_{\rm bar} - R_{\rm quadrupole}$, as obtained in sect.~\ref{sec:quadrupole_strength} (see Fig.~\ref{fig:bar_quadrupole_correlation_allmodels}) from a wide variety of simulated bar models, the inferred $S_{\rm bar}$ and $R_{\rm bar}$ would be $\sim 0.57$  and $\sim 4.5 \kpc$, respectively. Furthermore, the computed $\varphi_4$ for the selected sample of RGB stars turns out to be $\sim 5 \degrees$, thereby implying the MW's bar will be at an angle of $\sim 5 \degrees$ with respect to the Sun (for details, see sect.~\ref{sec:quadrupole_orientation}). At this point, it is tempting to estimate the `true' bar strength and length for the MW, given how the \gaia-like mock data overestimates these values by 35-45 percent, as shown previously. However, for a rigorous inference, first we need to check the universality and robustness of the \gaia-like mocks produced from different high-resolution simulated MW-like galaxies. This is beyond the scope of this work, and will be taken up in a future work.
\par

\section{Summary and future direction}
\label{sec:conclusion}

In summary, we investigated the formation and temporal evolution of the quadrupole feature in the stellar mean radial velocity field, excited by an $m=2$ stellar bar. In addition, we carried out a thorough study to test whether this quadrupole feature can be used as a robust kinematic diagnostic to put stringent constraints on the bar properties. To achieve that, we used a suite of 14 isolated, collisionless $N$-body models (having both thin and thick stellar discs) as well as a sample of barred galaxies from the TNG50 cosmological simulation. Finally, we investigated the possibility of constraining the MW's bar properties from the quadrupole feature in the stellar $\avg{V_R}$ field. Our main findings are listed below.\\

In absence of observational errors: 
\begin{itemize}
\item{A prominent stellar bar always excites a `butterfly' or quadrupole structure in the stellar mean radial velocity $\avg{V_R}$ field. We further devised a quantitative method to measure the strength ($S_{\rm quadrupole}$) and length ($R_{\rm quadrupole}$) via Fourier decomposition of the $\avg{V_R}$ distribution. The strength and the length of the quadrupole are strongly correlated with the strength and length of the bar, irrespective of the choice of thin or thick disc stars. In addition, the orientation of the quadrupole in the stellar kinematics (calculated from the $m=4$ Fourier phase angle) robustly captures the bar orientation in the density distribution. These trends hold true for all barred models (isolated or from the TNG50) used in this work. }

\item{ The strengths of the bar and the quadrupole tend to follow an empirical linear scaling relation of the form $Y = AX+B$ ($A = 0.96 \pm 0.01$; $B = -0.18 \pm 0.005$). Similarly, the lengths of the bar and the quadrupole tend to follow an empirical linear scaling relation of the form $Y = AX+B$ ($A = 1.06 \pm 0.01$; $B = 0.01 \pm 0.01$), However, the best-fit values of these linear relations change (within $\sim 10-25$ percent) depending on the choice of thin or thick disc stars.}

\item{The presence of transient spirals systematically overestimates the length of the quadrupole feature (with a median relative difference of $\sim 25$ percent). In such a bar+spiral dynamical scenario, like the MW, the constancy of $\varphi_4$ (within $\sim 5-8 \degrees$) serves as a much robust proxy for measuring the length of the quadrupole, and, in turn, inferring the bar length. }
\end{itemize}

In presence of observational errors:

\begin{itemize}

\item{In \gaia-like mock datasets, constructed from the simulated model while incorporating the dust extinction and the broad trends of \gaia\ -like uncertainties and systematic errors, the quadrupole properties are overestimated by $\sim 35-45$ percent when compared with their `true' values. We demonstrate that the majority of this effect comes from the uncertainty in parallax measurement from the \gaia\ survey.}

\end{itemize}

To conclude, we demonstrate that the quadrupole feature in the $\avg{V_R}$ distribution is indeed an excellent kinematic diagnostic to put stringent constraint on the bar properties (strength, length, and orientation), provided there are no significant observational errors involved. It will be worthy checking whether the scaling relations, derived in this work, between the length and strength of the bar and the quadrupole holds true for a diverse bar models. Furthermore, we caution that inferring MW's bar properties by using the stellar kinematic information from the \gaia\ DR3 when proper care is not taken into account for the observational errors (predominantly the uncertainty in parallax measurement) of the \gaia\ survey, can result in misleading conclusions about the MW's dynamics. We point out that this effect can be reduced by considering stars with more stringent cuts on parallax error. However, this comes at a cost from the number of available stars near the Galactic centre. With future data releases of \gaia\ DR4, this scenario is expected to improve. Alternatively, combining the \gaia\ parallax estimates with photometric surveys, for example, in \cite{Zhang24} can also improve this scenario.

\begin{acknowledgements}
We thank the anonymous referee for useful comments which helped to improve this paper. S.G. acknowledges funding from the IIT-Indore, through a Young Faculty Research Seed Grant (project: `INSIGHT'; IITI/YFRSG/2024-25/Phase-VII/02). T.K., S.G., and G.M.M. acknowledge funding from the Alexander von Humboldt Foundation, through a Sofja Kovalevskaja Award. P.D.M., D.K., and M.H. acknowledge the support of the French Agence Nationale de la Recherche (ANR), under grant ANR-13-BS01-0005
(project ANR-20-CE31-0004-01 MWDisc) This work has made use of the computational resources obtained through the DARI grant A0120410154 (P.I. : P. Di Matteo).
\end{acknowledgements}

\bibliographystyle{aa.bst} 
\bibliography{my_ref.bib} 

\begin{appendix}

\section{Temporal evolution of quadrupole properties}
\label{appen:density_and_Vr_maps}

\begin{figure*}
\centering
\resizebox{1.0\linewidth}{!}{\includegraphics{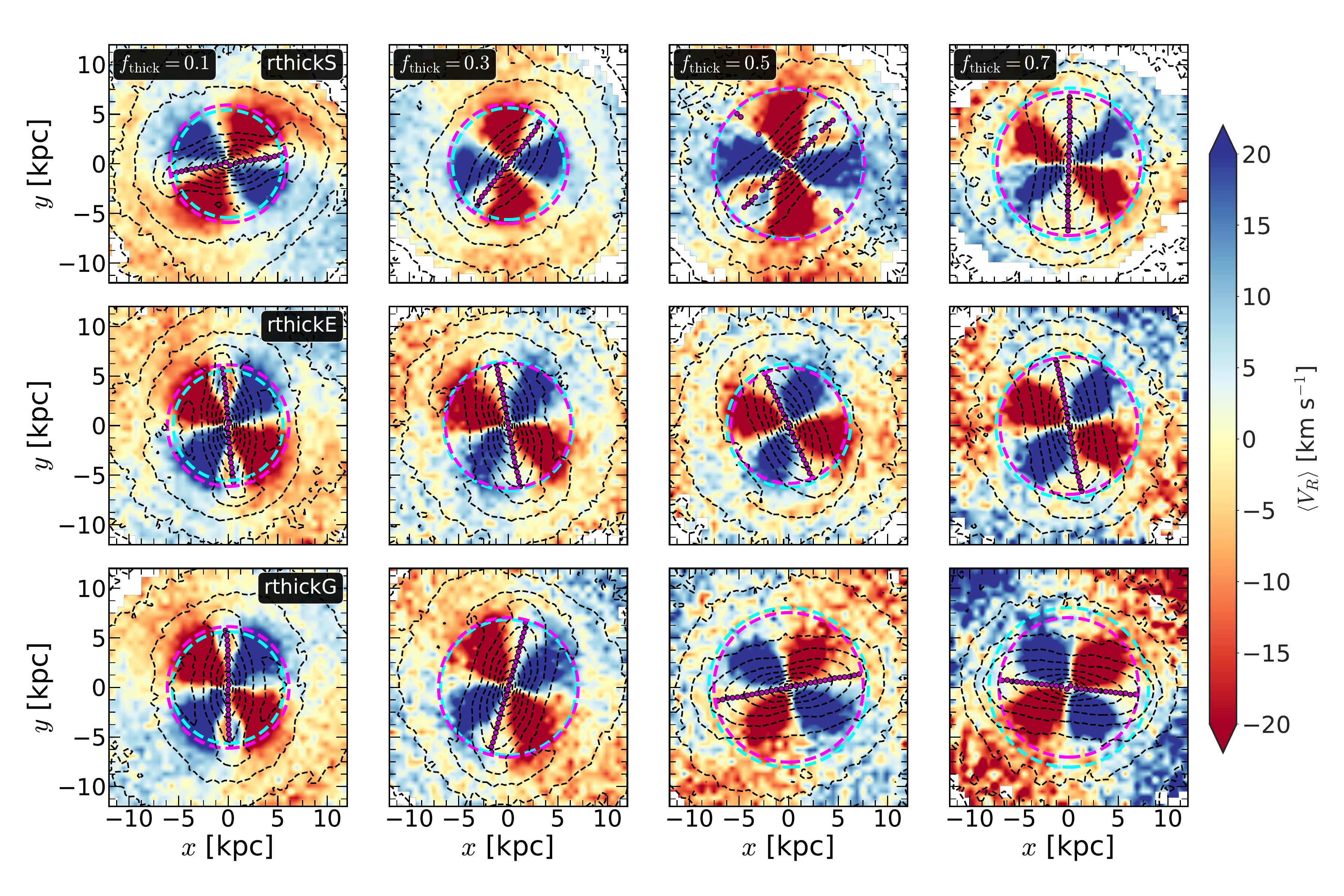}}
\caption{Distribution of the mean radial velocity, $\avg{V_R}$ in the $(x-y)$-plane, calculated at the end of the simulation run ($t = 9 \Gyr$), for all thin+thick models considered here. Black dashed lines denote the contours of constant surface density. The cyan dashed circle denotes the bar length, $R_{\rm bar}$, and the magenta dashed circle denotes the extent of the quadrupole feature, $R_{\rm quadrupole}$. The magenta points denote the spatial distribution of the phase-angle of the $m=4$ Fourier moment ($\varphi_4$). \textit{Top row} corresponds to the rthickS models whereas \textit{middle} and \textit{bottom row} correspond to rthickE and rthickG models, respectively. The thick disc mass fraction ($f_{\rm thick}$) varies from 0.1 to 0.7 (from left to right panels). A prominent quadrupole feature is present in all thin+thick models considered here.}
\label{fig:Vr_maps_allmodels_endstep}
\end{figure*}

\begin{figure*}
\centering
\resizebox{1.0\linewidth}{!}{\includegraphics{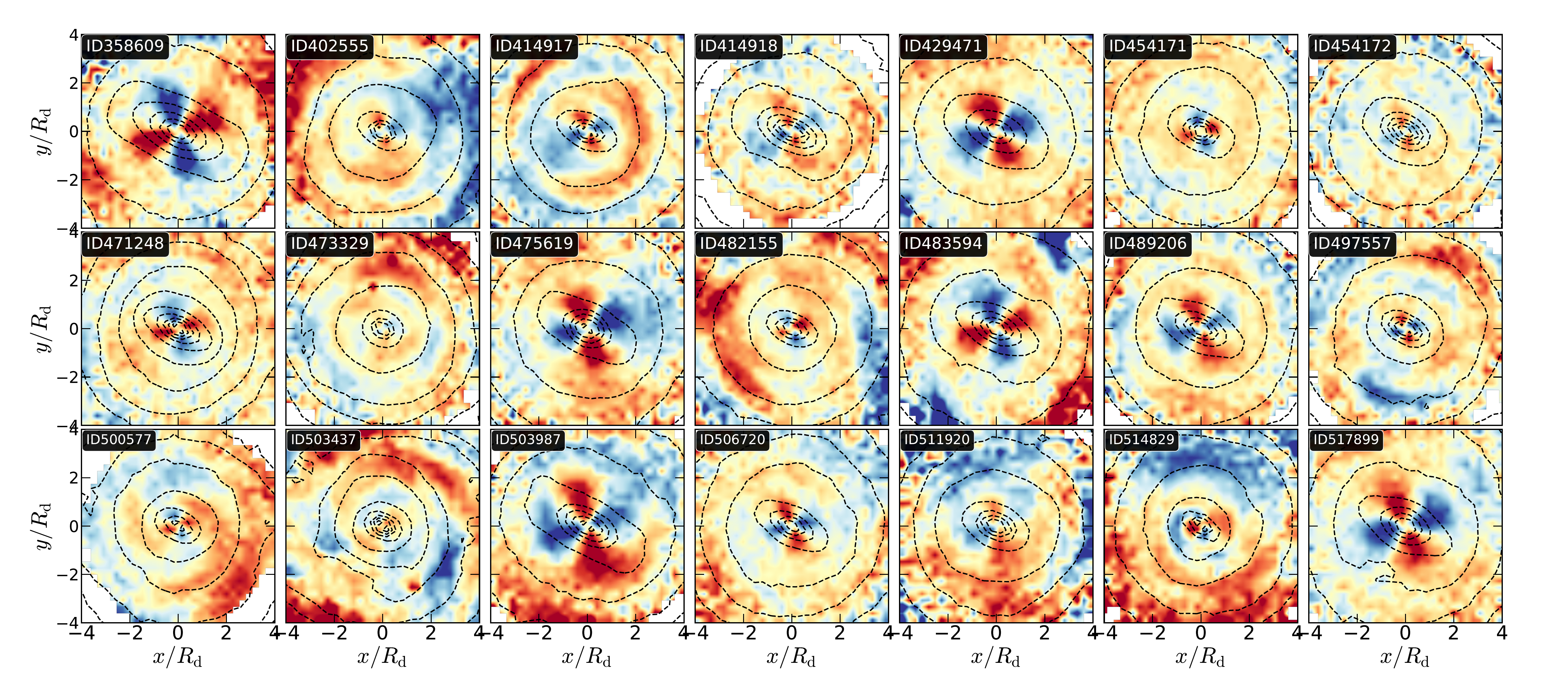}}
\caption{Distribution of the mean radial velocity, $\avg{V_R}$ in the $(x-y)$-plane, for some of the barred galaxies from the TNG50 simulations, considered in this work. For details, see the text. Black dashed lines denote the contours of constant surface density. A prominent bar is always accompanied by a prominent quadrupole feature in the $\avg{V_R}$ distribution. The colour bar is same as in Fig.~\ref{fig:Vr_maps_allmodels_endstep}. }
\label{fig:Vr_maps_TNG50_endstep}
\end{figure*}

\begin{figure*}
\centering
\resizebox{0.85\linewidth}{!}{\includegraphics{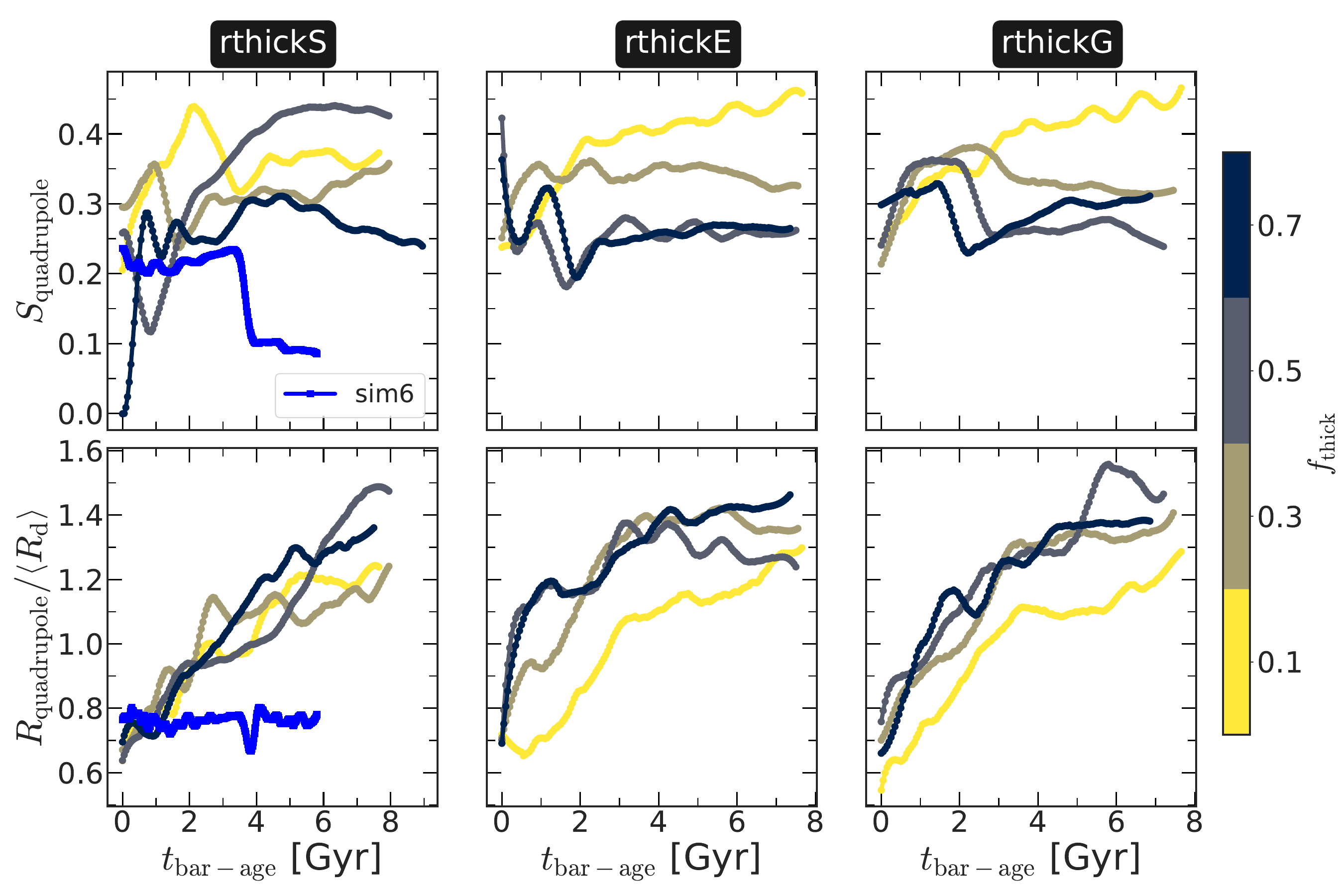}}
\caption{Temporal evolution of the strength of the quadrupole, $S_{\rm quadrupole}$ and the extent of the quadrupole, $R_{\rm quadrupole}$ (normalised by the mean disc scale length, $\avg{R_{\rm d}}$), for thin+thick models with different $f_{\rm thick}$ values (see the colour bar), as a function of bar age ($t_{\rm bar-age}$). Left panels correspond to the rthickS models whereas middle and \textit{right panels} correspond to rthickE and rthickG models, respectively. The strength and extent of the quadrupole for the model sim6 are shown in blue lines (see the left panels).}
\label{fig:Strength_quadrupole_temporal_allmodels_endstep}
\end{figure*}

Fig.~\ref{fig:Vr_maps_allmodels_endstep} shows the distribution of the stellar mean radial velocity ($\avg{V_R}$) in the $(x-y)$-plane, calculated at $t = 9 \Gyr$, for all 13 thin+thick models with varying $f_{\rm thick}$ values. A prominent bar is always accompanied by a prominent quadrupole feature in the $\avg{V_R}$ distribution. In each case, the extents of the bar and the quadrupole feature as well as the orientation of the quadrupole feature are indicated. In all cases, the orientation of the quadrupole feature traces accurately the bar orientation, thereby demonstrating that the quadrupole feature is a robust kinematic tracer of bar properties. Similarly, in Fig.~\ref{fig:Vr_maps_TNG50_endstep}, we show the corresponding distribution of the $\avg{V_R}$ in the $(x-y)$-plane, for a sample of TNG50 MW-like barred galaxies. The quadrupole feature traces accurately the bar orientation for these TNG50 galaxies as well. In addition, Fig.~\ref{fig:Strength_quadrupole_temporal_allmodels_endstep} shows the temporal evolution of the strength and extent of the quadrupole (for definitions, see sect.~\ref{sec:quadrupole_strength}) for all 14 isolated thin+thick bar models. 

\section{Further comparison of Gaia-like mock datasets}
\label{appen:more_on_GaialikeMocks}

\begin{figure}
\centering
\resizebox{\linewidth}{!}{\includegraphics{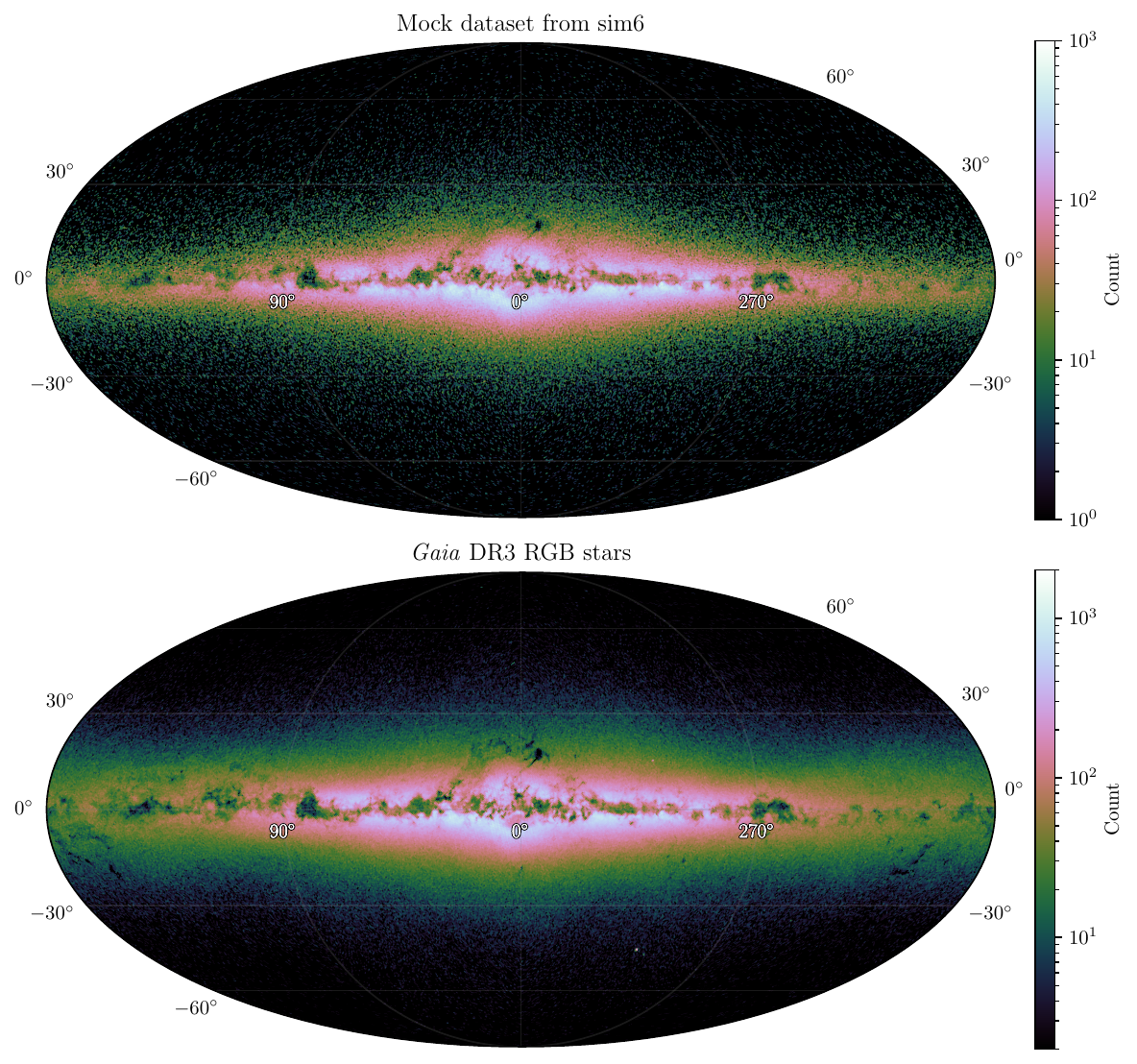}}
\caption{On-sky map in Galactic coordinates, for the \gaia-like mock dataset created from the bar model sim6 (\textit{top panel}) and our selected sample of \gaia\ DR3 stars (bottom panel). }
\label{fig:more_on_GaialikeMock}
\end{figure}

Fig.~\ref{fig:more_on_GaialikeMock} shows the on-sky distribution of stars in Galactic coordinates for the \gaia-like mock dataset created from the bar model sim6 and our selected sample of \gaia\ DR3 stars. Even a visual inspection reveals that the final mock dataset has a spatial distribution that closely mimics that of the \gaia\ DR3 sample once we incorporate the dust extinction, \gaia\ selection function and broad trends of \gaia\ -like uncertainties and systematic errors.

\end{appendix}

\end{document}